\documentclass[epj]{svjour}
\usepackage{graphicx}
\usepackage{cite}
\newcommand{\as}  {\ensuremath{\alpha_{\mathrm{S}}}}

\newcommand{\asmz}{\ensuremath{\alpha_\mathrm{S}(M_{\mathrm{Z^0}})}}
\newcommand{\Jade}{\mbox{\rm JADE}}
\newcommand{\JadeBF}{\mbox{\rm\bf JADE}}
\newcommand{\Opal}{\mbox{\rm OPAL}}
\newcommand{\OpalBF}{\mbox{\rm\bf OPAL}}
\newcommand{\epem}{\ensuremath{\mathrm{e^+e^-}}}

\newcommand{\stat}{\ensuremath{\mathrm{(stat.)}}}

\newcommand{\expt}{\ensuremath{\mathrm{(exp.)}}}
\newcommand{\had} {\ensuremath{\mathrm{(had.)}}}
\newcommand{\theo}{\ensuremath{\mathrm{(theo.)}}}
\def\resJ{\ensuremath{
  \asmz=0.1287\pm0.0007\stat\pm0.0011\expt\pm0.0022\had\pm0.0075\theo}}
\def\resJlines{
  \begin{eqnarray*}
  \asmz&=&0.1287\pm0.0007\stat\pm0.0011\expt\nonumber\\
       & &\pm0.0022\had\pm0.0075\theo\,,
  \end{eqnarray*}
}
\newcommand{\ytwothree}   {{\ensuremath{y^{\rm D}_{23}}}} 
\newcommand{\tma}               {\ensuremath{T_{\mathrm{maj}}}}
\newcommand{\tmi}               {\ensuremath{T_{\mathrm{min}}}}
\newcommand{\rs}                {\ensuremath{\sqrt{s}}}
\newcommand{\cp}                {\ensuremath{C}}
\newcommand{\bt}                {\ensuremath{B_\mathrm{T}}}
\newcommand{\bw}                {\ensuremath{B_\mathrm{W}}}
\newcommand{\bn}                {\ensuremath{B_\mathrm{N}}}
\newcommand{\mh}                {\ensuremath{M_\mathrm{H}}}
\newcommand{\ml}                {\ensuremath{M_\mathrm{L}}}
\newcommand{\thr}               {\ensuremath{1-T}}
\newcommand{\momn}[2] {\mbox{\ensuremath{\langle#1^{#2}\rangle}}}
\newcommand{\dd}       {\mathrm{d}}
\newcommand{\zzero}     {\ensuremath{\mathrm{Z^0}}}
\newcommand{\asb}               {\ensuremath{\bar{\alpha}_{\mathrm{S}}}}
\newcommand{\asbsq}             {\ensuremath{\bar{\alpha}_{\mathrm{S}}^2}}
\newcommand{\xmu}               {\ensuremath{x_{\mu}}}
\newcommand{\nf}                {\ensuremath{n_{\mathrm{f}}}}
\newcommand{\sigtot} {\ensuremath{\sigma_{\mathrm{tot}}}}
\newcommand{\signull}{\ensuremath{\sigma_{0}}}
\newcommand{\invpb}     {\ensuremath{\mathrm{pb}^{-1}}}
\newcommand{\py}                {PYTHIA}
\newcommand{\hw}                {HERWIG}
\newcommand{\ar}                {ARIADNE}

\newcommand{\bbbar}     {\ensuremath{\mathrm{b\bar{b}}}}
\newcommand{\rsp}               {\ensuremath{\sqrt{s'}}}
\newcommand{\cdet}    {\ensuremath{C_{\mathrm{det}}}}

\newcommand{\oaa}               {\ensuremath{\mathcal{O}(\alpha_\mathrm{S}^2)}}
\newcommand{\chad}    {\ensuremath{C^{\mathrm{had}}}}
\def\be{\begin{equation}}
\def\ee{\end{equation}}
\newcommand{\chisq}     {\ensuremath{\chi^2}}
\newcommand{\mz}                {\ensuremath{M_{\mathrm{Z^0}}}}
\newcommand{\nlo}{\mbox{NLO}}
\newcommand{\chisqd}    {\ensuremath{\chi^2/\mathrm{d.o.f.}}}
\newcommand{\momone}[1] {\mbox{\ensuremath{\langle#1\rangle}}}
\newcommand{\asi}                {\ensuremath{\alpha_{\mathrm{S},i}}}
\def\resJOlines{\begin{eqnarray*}
  \asmz&=&0.1262\pm0.0006\stat\pm0.0010\expt\nonumber\\
       & &\pm0.0007\had\pm0.0064\theo\,,  \end{eqnarray*}}
\newcommand{\resJtot}  {\ensuremath{\asmz=0.1287\pm0.0079~(\mathrm{total~error})}}
\newcommand{\resJOtot} {\ensuremath{\asmz=0.1262\pm0.0065~(\mathrm{total~error})}}
\newcommand{\Petra} {\mbox{\rm PETRA}}
\newcommand{\Herwig}{\mbox{HERWIG}}
\newcommand{\Ariadne}{\mbox{ARIADNE}}
\newcommand{\Lep}{\mbox{LEP}}
\newcommand{\lo}{\mbox{LO}}

\begin{document}
\titlerunning{Study of moments of event shapes and a determination of \as\ using \epem\ annihilation data from \Jade}
\title{       Study of moments of event shapes and a determination of \boldmath{\as} using \boldmath{\epem} annihilation data from \JadeBF}
\author{C. Pahl\inst{1,}\inst{2} \and S. Bethke\inst{1} \and  S. Kluth\inst{1} \and  J. Schieck\inst{1} 
        \and the \Jade\  Collaboration\inst{3}
}                     
%
%
\institute{Max-Planck-Institut f\"ur Physik, F\"ohringer Ring 6, D-80805 Munich, Germany
      \and Excellence Cluster Universe, Technische Universit\"at M\"unchen,
           Boltzmannstr. 2, D-85748 Garching, Germany
      \and See~\cite{naroska} for the full list of authors.}
\date{Received: date / Revised version: date\\
\\
In memory of Beate Naroska.}

\abstract{  Data from \epem\ annihilation into hadrons, collected by the \Jade\ 
  experiment at centre-of-mass energies between 14~GeV and 44~GeV,
  are used to study moments of event shape distributions. 
  Models with hadronisation parameters tuned to the LEP~1 precision data provide
  an adequate description of the low energy data studied here. The NLO QCD calculations,
  however, show systematic deficiencies for some of the moments.
  The strong coupling  measured from the moments which are reasonably described by NLO QCD, 
\begin{center}
  \resJ\,,
\end{center}
is consistent with the world average.
\PACS{
      {12.38.Bx}{Perturbative calculations}             \and
      {12.38.Qk}{Experimental tests} 
     } 
} 
\maketitle
\section{Introduction}
\label{intro}
Electron-positron annihilation into hadrons constitutes a precise 
testing ground of Quantum Chromodynamics\linebreak (QCD).  Commonly jet
production rates or distributions of event shape variables have been
studied.  Predictions of perturbative QCD combined with hadronisation
corrections derived from models have been found to describe the data
at low and high energies well, see
e.g.~\cite{jadenewas,OPALPR299,movilla02b,OHab,STKrev}.

In this analysis we use data from the \Jade\ experiment, recorded in the
years 1979 to 1986 at the PETRA \epem\ collider at DESY at six
centre-of-mass (c.m.) energies \rs\ covering the range 14--44~GeV. 
We
measure the first five moments of event shape variables for the
first time in this low \rs\ region of \epem\ annihilation and compare the
data to predictions by Monte Carlo (MC) models and by perturbative QCD.
Moments sample all phase space, but 
are more sensitive to specific parts of phase space, dependent on their order.
From the comparison of the data with theory we extract the
strong coupling \as. 
The measurement of the moments, as well as the \as~determination, follow closely 
the analysis by the \Opal\  experiment in the complete
LEP energy range of 91--209~GeV \cite{OPALPR404}. 
 This work supplements our previous analyses on jet production rates, determinations of \as\
 and four jet production, using \Jade\ and \Opal\ data \cite{OPALPR299,VierjetJADE,VierjetJO}.

The outline of the paper is as follows. In Sect.~\ref{theory}, we present the observables 
used in the analysis and describe the perturbative QCD predictions. 
In Sect.~\ref{analysis_procedure} the analysis procedure is
explained in detail. Sect.~\ref{systematic} contains the discussion of the
systematic checks which are performed and the resulting systematic
errors. We collect the results and describe the determination of \as\ in 
Sect.~\ref{results}, 
and we summarize in Sect.~\ref{summary}.

\section{Observables}
\label{theory}

Event shape variables are a convenient way to characterise properties of hadronic events
by the distribution of particle momenta. For the definition of the variables we refer to \cite{OPALPR404}. 
The event shapes considered here are
Thrust {T},
C-parameter {C},
Heavy Jet Mass {\mh}, 
jet broadening variables {\bt} and {\bw},
and the transition value between 2 and 3 jets in the Durham jet scheme, {\ytwothree}. 
The \as\ determination
in \cite{OPALPR404} is based on distributions and moments of these variables.
Their theo\-re\-ti\-cal description is currently the most advanced
\cite{resummation,NNLOESs,Weinzierl}. Further, we  
measure moments of Thrust major {\tma}, Thrust minor {\tmi}, Oblateness {O}, Spheri\-city {S}, 
Light Jet Mass {\ml}, and Narrow Jet Broadening {\bn}. Moments of these variables and 
variances of all measured event shapes will be made available in the HEPDATA database.\footnote{\tt http://durpdg.dur.ac.uk/HEPDATA/}.

Generic event shape variables $y$ are constructed such that spherical and multi-jet events
yield large values of $y$, while
two narrow back-to-back jets yield $y\simeq0$. Thrust $T$ is an exception to this rule. By using 
$y=\thr$ instead the condition is fulfilled for all event shapes. 

The $n$th, $n=1,2,\ldots$, moment of the distribution of the event
shape variable $y$ is defined by
\be\momn{y}{n}=\int_0^{y_{max}} y^n \frac{1}{\sigma}
\frac{\dd\sigma}{\dd y} \dd y \,,\ee
where $y_{max}$ is the
kinematically allowed upper limit of the variable $y$.  

Predictions have been calculated for the moments of event shapes. Their evolution
with c.m. energy allows direct tests of the predicted energy evolution
of the strong coupling \as. Furthermore we 
determine \asmz\ by evolving our measurements to the energy scale given by the mass of the \zzero\ boson.
The theoretical calculations
involve a integration over full phase space, which implies that
comparison with data always probes all of the available phase space.
This is in contrast to QCD predictions for distributions; these are
commonly only compared with data--e.g.\ in order to measure \as--in
restricted regions, where the theory is well defined and describes the data
well, see e.g.~\cite{jadenewas}.
Comparisons of QCD predictions for
moments of event shape distributions with data are thus complementary
to tests of the theory using distributions.

Uncertainties in the NNLO predictions for event shape distributions in the two-jet region~\cite{NNLOESs,Weinzierl} prevent the reliable calculation of moments to NNLO at present, and therefore we compare with NLO predictions only.
The QCD prediction of \momn{y}{n} at parton level, in next-to-leading order (NLO) perturbation theory, and with
$\asb\equiv\as/(2\pi)$, is
\begin{equation}
  \momn{y}{n}^{\rm part,theo} = {\cal A}_n \,\asb + ({\cal B}_n-2{\cal A}_n) \,\asbsq\,.
\label{eq_qcdmom} 
\end{equation}
The values of the coefficients\footnote{The \asbsq\ coefficient is written as ${\cal B}_n-2{\cal A}_n$
because the QCD calculations are normalized to the Born cross section $\sigma_0$,
while the data are normalized to the total hadronic cross section, 
$\sigtot=\signull(1+2\asb)$ in LO.} ${\cal A}_n$ 
and ${\cal B}_n$ can be obtained by numerical 
integration of the QCD matrix
elements using the program EVENT2~\cite{event2}.  

The coupling \asb\ and the \asbsq\ coefficient depend on the renormalisation scale $\mu$~\cite{ert}.  
For the sake of clarity the renormalisation scale factor is defined as
$\xmu\equiv\mu/\rs$, so setting \xmu\ to one implies that the
renormalisation scale is \rs.
A truncated fixed order QCD calculation such
as~(\ref{eq_qcdmom}) will then depend on \xmu.  The renormalisation
scale dependence is implemented by the replacement ${\cal
  B}_n\rightarrow {\cal B}_n+\beta_0\ln(\xmu){\cal A}_n$ where
$\beta_0=11-\frac{2}{3}\nf$ is the leading order $\beta$-function
coefficient of the renormalisation group equation and $\nf=5$ is the
number of active quark flavours.

\section{Analysis procedure}
\label{analysis_procedure}
\subsection{The \JadeBF\ detector}
\label{sec_detector}

The JADE detector is described in detail
in ref.~\cite{naroska}. 
Energy measurement by the electromagnetic calorimeter and
the reconstruction of charged particle tracks in the central track detector
are the main ingredients for
this analysis.
The central jet chamber was 
positioned in an axial magnetic field of 0.48 T
provided by a solenoidal magnet.\footnote{{In the JADE 
right-handed coordinate system the $+x$ axis pointed towards the
centre of the PETRA ring, the $y$ axis pointed upwards and the $z$
axis pointed in the direction of the positron beam. The polar angle
$\theta$ and the azimuthal angle $\phi$ were defined with respect to
$z$ and $x$, respectively, while $r$ was the distance from the
$z$-axis.}}  The magnet coil was surrounded by the lead glass calorimeter,
which measured electromagnetic energy and consisted of a 
barrel and two endcap sections.

\subsection{Data samples}
\label{data_samples}

In this analysis we are using data samples identical to the samples used in~\cite{VierjetJADE,naroska,jadenewas,OPALPR299,movilla02b,pedrophd},
collected by the JADE~experiment between 1979 and
1986; they correspond to a total integrated luminosity of ca. 195 \invpb.
Table~\ref{lumi} contains the breakdown of the data samples--data taking period, energy range, 
mean centre-of-mass energy,  
integrated luminosity and the number of selected hadronic events.  
\begin{table}[h]
\caption{Year of data taking, energy range, integrated luminosity,
average centre-of-mass energy and the numbers of selected data events 
for each data sample}
\label{lumi}
\begin{tabular}{ l l l l r }
\hline\noalign{\smallskip}
year       & range of     & \rs\ mean     & luminosity & selected \\
           & \rs\ in GeV  &        in GeV & (pb$^{-1}$)  & events   \\
\noalign{\smallskip}\hline\noalign{\smallskip}
1981       & 13.0--15.0   & 14.0          &  1.46      & 1783  
\\
1981       & 21.0--23.0   & 22.0          &  2.41      & 1403  
\\ 
1981--1982 & 33.8--36.0   & 34.6          & 61.7       & 14313 
\\
1986       & 34.0--36.0   & 35.0          & 92.3       & 20876 
\\ 
1985       & 37.3--39.3   & 38.3          &  8.28      & 1585  
\\ 
1984--1985 & 43.4--46.4   & 43.8          & 28.8       & 4376  \\ 
\noalign{\smallskip}\hline
\end{tabular}
\end{table}

\subsection{Monte Carlo samples}
\label{MCsamples}

To correct the data for experimental effects and backgrounds
we use samples of MC simulated events. 
Using \py~5.7~\cite{jetset3} we simulate
the process $\epem\to\mathrm{hadrons}$.
For systematic checks we use
corresponding samples obtained by simulation with \hw~5.9~\cite{herwig51}.
We process the MC samples generated at each energy point
through a full si\-mu\-la\-tion of the JADE detector~\cite{jadesim1,jadesim2,jadesim3}, summarized in ~\cite{pedrophd};
and we reconstruct them in essentially the same way as the data.
 
Using the parton shower models \py~6.158, \hw~6.2~\cite{herwig6} and \ar~4.11~\cite{ariadne3}
we employ in addition large samples of
MC events without detector
simulation, in order to compare with the
corrected data. For the purpose of
comparison with the data, the MC
events include the effects of
hadronisation, i.e. the transition of
partons into hadrons.
All used major versions of the models were adjusted to LEP~1 data by the OPAL 
collaboration~\cite{OPALPR141,OPALPR379}, so we expect comparable results
from them.

\subsection{Selection of events}
\label{selection}

The selection--identical to the one used in~\cite{VierjetJADE}--aims at selecting hadronic
events in the JADE data
excluding events with much energy
lost by initial state radiation (ISR). 
The rejected background consists to a large degree of two photon events.
It uses cuts on event multiplicity,
on visible energy and longitudinal momentum 
balance. 
The cuts are documented in~\cite{StdEvSel1,StdEvSel2,StdEvSel3,jadenewas}.

So called good tracks and calorimeter clusters are identified by appropriate
criteria \cite{VierjetJADE}.
Double counting of energy from charged tracks and calorimeter
clusters is avoi\-ded by subtracting the estimated contribution
of a charged track from the associated cluster energy. 

The number of selected events for each energy point is given in
table~\ref{lumi}.

\subsection{Corrections to the Data}
\label{detectorcorrection}

The data are corrected further for the effects of limited detector acceptance and resolution,
and residual ISR following \cite{VierjetJADE}. 
All selected charged tracks, as well as the electromagnetic
calorimeter clusters remaining after the correction for
double counting of energy as described above, are used 
in the evaluation of the event shape moments.
The values of the moments after the application of all selection cuts are said 
to be at the detector level.

As the QCD predictions are calculated for massless quarks
we have to correct our data for the presence of events originating from \bbbar\ final states.
Especially at low \rs\ the large
mass of the b quarks and of the subsequently produced and decaying B hadrons will
influence the values of the event shape variables.
Therefore
in the \Jade\ analysis events from the process $\epem\rightarrow\bbbar$ 
(approximately 1/11 of the hadronic events) are
considered as background.

For the determination of the moments we calculate the sums $\sum_i
y_{i,\rm data}^n$ (for moment order $n=1,\ldots,5$) where $i$ counts all selected events.
The expected contribution of \bbbar\ background events $\sum_i
y^n_{i,\bbbar}$, as estimated by \py, is subtracted from the observed sum $\sum_i y^n_{i,\rm data}$.
By a multiplicative correction we then account for   
the effects of detector imperfections and
of residual ISR and two photon background.

Two sets of sums $\sum_i y^n_i$ are calculated from MC simulated
signal events.
At detector level, MC events are treated
identically to the data. The 
hadron level set is computed using the true momenta of the stable
particles in the event\footnote{ All charged and neutral particles
with a lifetime larger than $3\times 10^{-10}$~s are considered
stable.}, and uses only events where $\rsp$, the c.m.
energy of the event, reduced due to ISR, satisfies $\rs-\rsp<0.15$~GeV.
The 
ratio of the MC hadron level moment over the MC detector level moment 
is applied as a detector correction factor for the data; the corrected sums are
normalized by the corrected total number of selected events $N_{\rm tot}$ yielding the
final values of \momn{y}{n},
  \begin{equation}
  \momn{y}{n}^{\rm had} = \frac{\momn{y}{n}^{\rm had,MC}}{\momn{y}{n}^{\rm det,MC}} 
  \cdot \left( \sum_i{y}^{n}_{i,\rm det} - \sum_i{y}^{n}_{i,\rm \bbbar} \right)/N_{\rm tot}\,.
  \label{detcor-eq}
  \end{equation}
The corrected total number of events is
calculated from the number of selected events in the data in the same
way as for the moments.

\begin{figure}
\begin{center}
\includegraphics[width=0.47\textwidth]{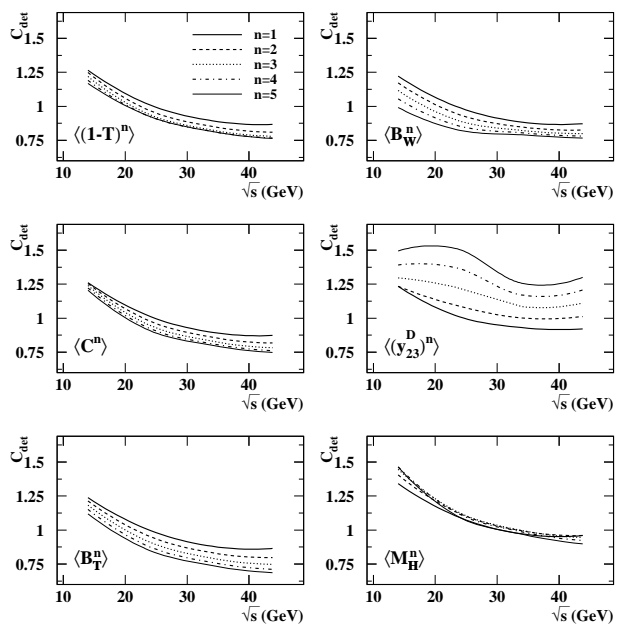}
\end{center}
\caption{Detector correction factors 
  $C_{\rm det}$ as calculated 
  using the \py\ MC model (see text for details). Line types correspond to
  moment order as shown in {\it top left} figure}
\label{detcor}
\end{figure}

There 
is some disagreement between the detector corrections calculated using 
\py\ or \hw\ at low \rs\ while at larger \rs\ the correction factors
agree well for most observables. The difference in detector corrections 
will be evaluated as an experimental systematic uncertainty, see Sect.~\ref{systematic}.
The detector correction factors\linebreak \cdet$={\momn{y}{n}^{\rm had,MC}}/{\momn{y}{n}^{\rm det,MC}}$ as determined using 
PY\-THIA are shown in Fig.~\ref{detcor}. 

\section{Systematic uncertainties}
\label{systematic}

Several contributions
to the experimental uncertainties are estimated by repeating the
analysis with varied track or event selection cuts or varied procedures
as in~\cite{VierjetJADE}. For each systematic
variation the value of the event shape moment or of \as\ is determined 
and then compared to the default value. 
The experimental systematic uncertainty 
quoted is the result of adding in quadrature all contributions. 
In the fits of
the QCD predictions to the data two further systematic uncertainties
are evaluated:

\begin{itemize}
\item Using \hw~6.2\ and
  \ar~4.11\ instead of \py~6.158\ we assess the uncertainties associated with the hadronisation 
  correction (Sect.~\ref{fitprocedure}). 
  The hadronisation systematic
  uncertainty is defined by the larger change in \as\ resulting from these
  alternatives.
\item By varying
  the renormalisation scale factor \xmu\ we assess the theoretical uncertainty associated 
  with missing higher
  order terms in the theoretical prediction.  
  The renormalisation scale factor \xmu\ is set to 2.0
  and 0.5.  
  The theoretical systematic uncertainty is defined by the larger deviation from the default 
  value.
\end{itemize}

\section{Results}
\label{results}
\subsection{Values of event shape moments}
\label{momentvalues}
\begin{figure*}
\begin{center}
\includegraphics[width=0.8\textwidth]{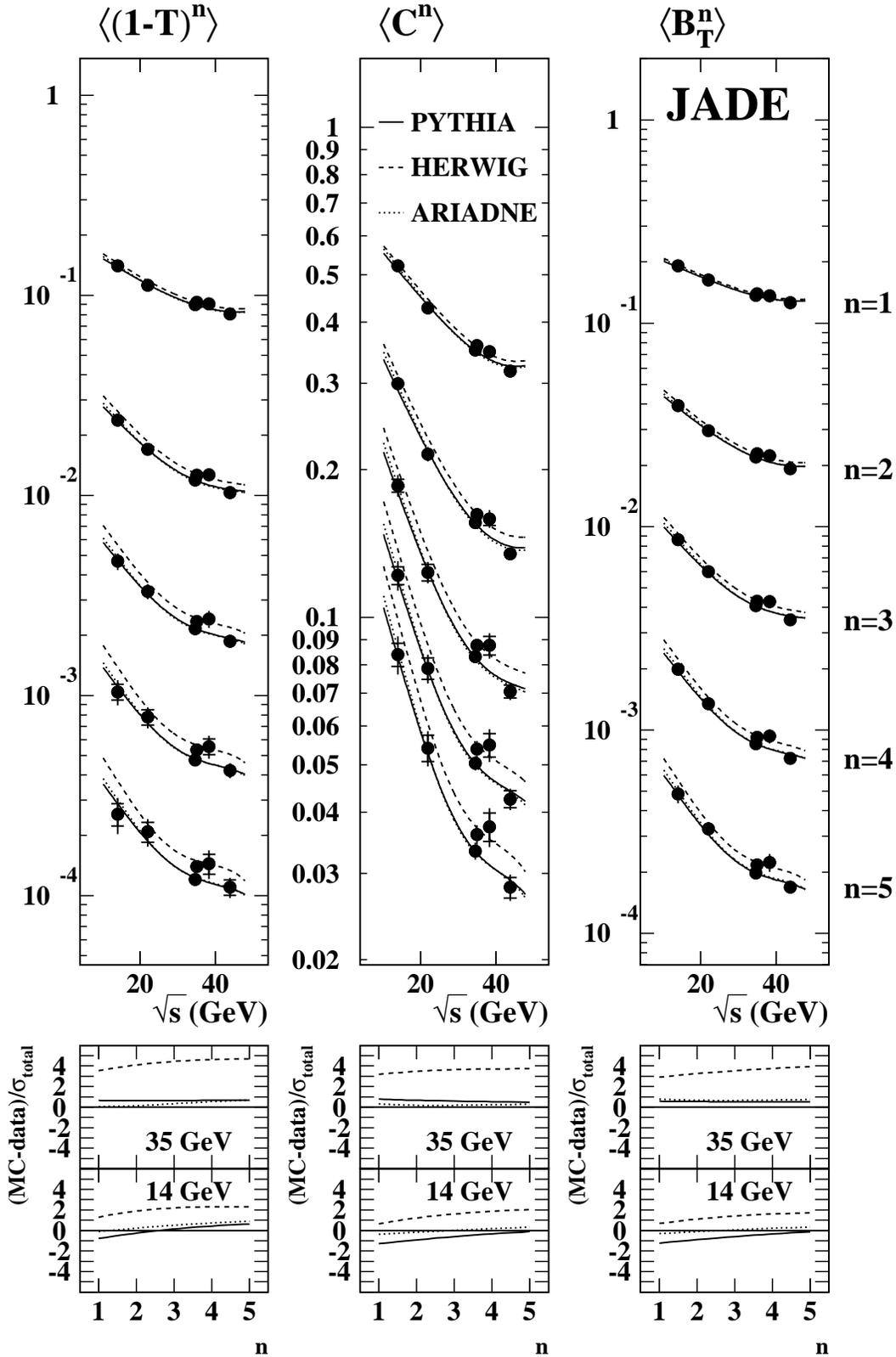} 
\end{center}
\caption{First five moments of \thr, \cp\ and \bt\ at hadron level 
  compared with predictions based on \py~6.158, \hw~6.2 and
  \ar~4.11 MC events.  
  The inner error bars--where visible--show the statistical errors,
  the outer bars show the total errors. Where no error bar is visible,
  the total error is smaller than the point size.
  The lower panels show the
  differences between data and MC at $\rs=14$~ and 35~GeV,
  divided by the total error}
\label{hadron}
\end{figure*}

\begin{figure*}
\begin{center}
\includegraphics[width=0.8\textwidth]{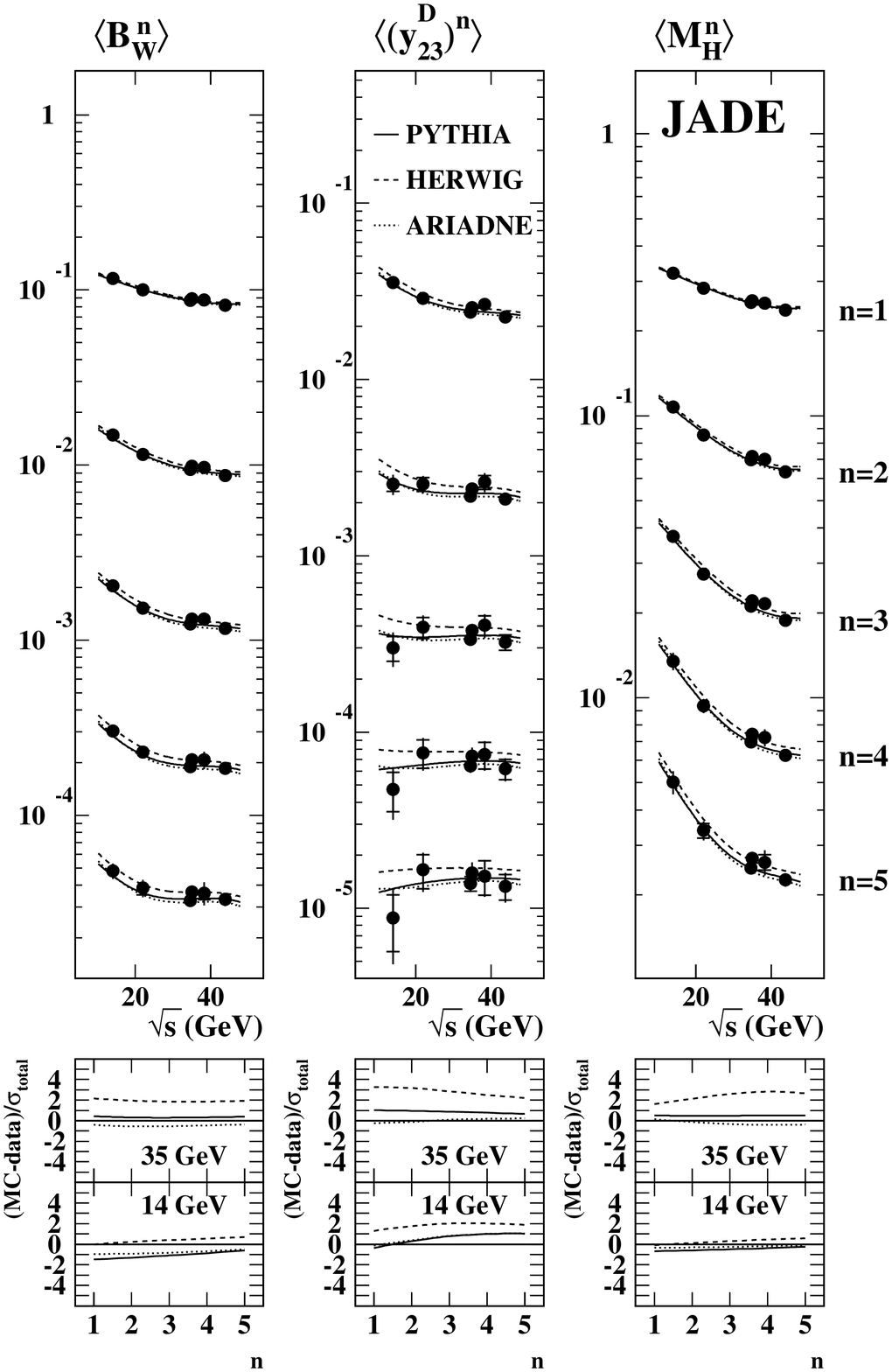}
\end{center}
\caption{First five moments of \bw, \ytwothree\ and \mh\ at hadron level 
  compared with predictions based on \py~6.158, \hw~6.2 and
  \ar~4.11 MC events.  
  The inner error bars--where visible--show the statistical errors,
  the outer bars show the total errors. Where no error bar is visible,
  the total error is smaller than the point size.
  The lower panels show the
  differences between data and MC at $\rs=14$~ and 35~GeV,
  divided by the total error}
\label{hadron2}
\end{figure*}

The first five moments of the six event shape variables after
subtraction of \bbbar\ background and correction for detector effects measured by \Jade\ are
listed in Tables~\ref{tabmomA} and \ref{tabmomB} and shown in Figs.~\ref{hadron} and 
\ref{hadron2}.  Superimposed we show
the moments predicted by the \py, \hw\ and \ar\ MC
models tuned by \Opal\ to LEP~1 data. The moments become smaller by approximately one order of
magnitude with each increasing moment order; the higher moments 
are more strongly suppressed
with centre-of-mass  
energy. Statistical and experimental systematic uncertainties strongly increase with moment
order.  

In order to make a clearer
comparison between data and models the lower plots in Figs.~\ref{hadron} and \ref{hadron2}  show the
differences between data and each model divided by the combined
statistical and experimental error for $\rs=14$ and 35~GeV.  The three
models are seen to describe the data fairly well; \py\ and \ar\ are
found to agree better with the data than \hw.

\subsection{Determination of \boldmath{\as}}
\label{fitprocedure}
In order to measure the strong coupling  \as,
we
fit the QCD predictions to the corrected moment values \momn{y}{n}, i.e. to the data
shown in Tables~\ref{tabmomA} and \ref{tabmomB}.  
The theoretical predictions using the \oaa\ calculation described in Sect.~\ref{theory} provide 
values at the parton level.  
It is necessary to correct for hadronisation effects
in order to compare the theory with the hadron level data. 
Therefore the moments are calculated at
hadron and parton level using large samples of \py\ 6.158 events and, as a 
cross check,
samples obtained by simulation with \hw~6.2 and \ar~4.11. 
Parton level is the stage at the end of the parton shower in the simulation of an hadronic event.
In order to correct for hadronisation
the data points are then multiplied by the ratio $\chad={\momn{y}{n}^{\rm part,MC}}/{\momn{y}{n}^{\rm had,MC}}$ of the parton over hadron 
level moments; $\momn{y}{n}^{\rm part}=\chad\cdot\momn{y}{n}^{\rm had}$.

The models use cuts on quantities like e.g. the invariant mass between
partons in order to regulate divergencies in the predictions for the
parton shower evolution.  As a consequence in some events no parton
shower is simulated and the original quark-antiquark pair enters the
hadronisation stage of the model directly. This leads to a bias in
the calculation of moments at the parton level, since $y=0$ in this
case for all observables considered here (\ytwothree\ cannot be calculated in this 
case).
In order to avoid this bias
we exclude in the simulation at the parton level events without
gluon radiation, as in \cite{Daisuke}.  
After this exclusion, the \rs\ evolution of the moments follows the QCD prediction;
the change of the prediction is comparable in size with the differences 
between employed MC generators.
At the hadron  and detector level all
events are used.

\begin{figure}
\includegraphics[width=0.47\textwidth]{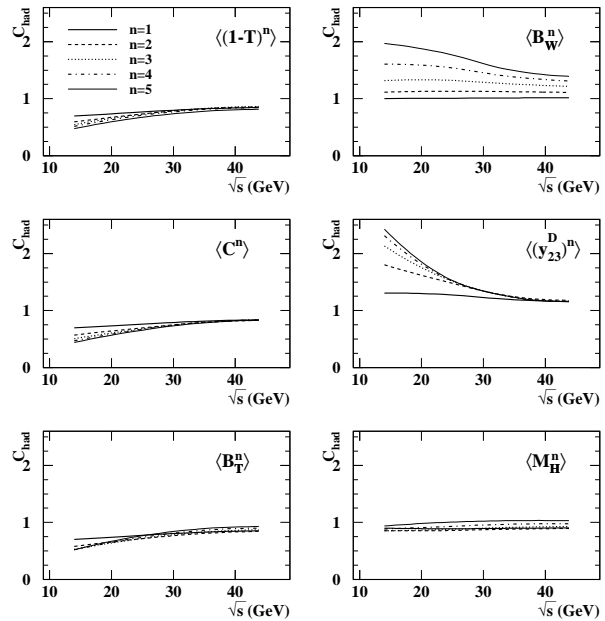}
\caption{Hadronisation correction factors $C^{\rm had}$
  as calculated using the MC model \py~6.158 (see text for details). Line types correspond to
  moment order as shown in {\it top left} figure}
\label{hadcor}
\end{figure}

The hadronisation correction factors \chad\ as obtained from \py~6.158
are shown in Fig.~\ref{hadcor}.  We find
that the hadronisation correction factors 
can be as large as two at low \rs.  For larger \rs\ the hadronisation corrections decrease as
expected. 

The models \py~6.158, \hw~6.2 and \ar\ 4.11 do not agree well for
moments based on \bw, \ytwothree\ and \mh\ at low \rs.  The
differences between the models are studied as a systematic
uncertainty in the fits.

A \chisq\ value for each moment \momn{y}{n} is calculated using the
formula
\be
  \chi^{2} = \sum_i (\momn{y}{n}^{\rm part}_i-\momn{y}{n}^\mathrm{part,theo}_i)^{2}/\sigma_{i}^{2}\,,
  \label{simplechi2}
\ee 
where $i$ counts the energy points, $\sigma_i$ denotes the 
statistical errors and $\momn{y}{n}^{\rm part,theo}$ is taken from~(\ref{eq_qcdmom}). 

The \chisq\ value is
minimized with respect to \asmz\ for each moment $n$ separately.  
The statistical uncertainty is found by varying the minimum value $\chisq_{\rm min}$ to $\chisq_{\rm min}+1$.
The
evolution of \as\ from \mz\ to c.m. energy $(\rs)_i$ is implemented in the fit in two-loop 
precision \cite{ESW}.
The renormalisation scale factor \xmu, as
discussed in Sect.~\ref{theory}, is set to~1.

 \subsection{Fits of \JadeBF\ data}
\label{JadeFits}
\begin{figure}[h]
\includegraphics[width=0.48\textwidth]{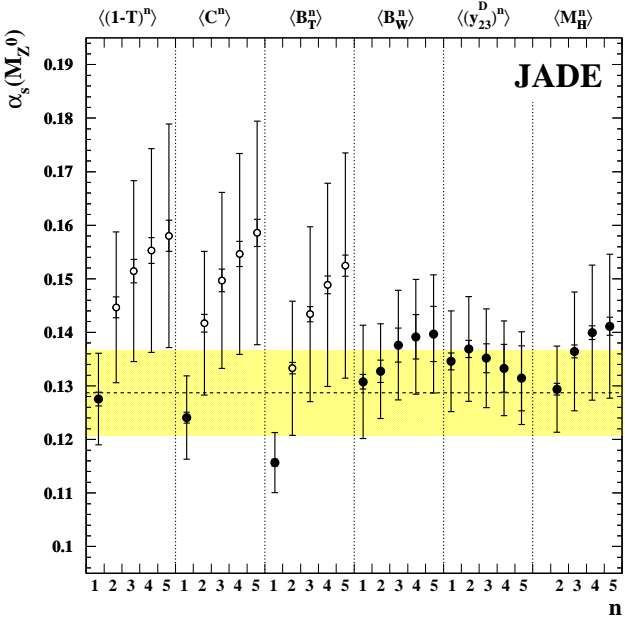} 
\caption{Measurements of \asmz\ using fits to moments of six event
  shape variables at \Petra\ energies.  
  The inner error bars--where visible--show the statistical errors,
  the outer bars show the total errors.
  The {\it dotted line} indicates
  the weighted average described in Subsect.~\ref{ascombs}, the {\it shaded band} shows its error.
  Only the measurements
  indicated by {\it solid symbols} are used for this purpose}
\label{fit_plot}
\end{figure}

  Data and \nlo\ prediction are compared, and this
  is repeated for every systematic variation.
  The results are\linebreak shown in Fig.~\ref{fit_plot} and listed in Table~\ref{JADE-note-dt}.
  Figure \ref{fit_plot} also contains the 
  combination of the fit results discussed below.
  The values of \chisqd\ are in the order of 1-10, the fitted predictions--including the
  energy evolution of \as--are consistent with the data.
  The fit to \momn{\mh}{1} does not converge and therefore no result is shown.\footnote{Equation \ref{eq_qcdmom} precludes a real solution \asb, if 
  ${\cal B}_n - 2 {\cal A}_n < -{\cal A}_n^2/4\momn{y}{n}$. For \momn{\mh}{1} this relation 
  is fulfilled in the whole energy range 14--207~GeV, see Tables~\ref{tabmomA} and \ref{tabmomB} 
and \cite{OPALPR404}. The \nlo\ coefficient 
is negative in the case of \momn{\bw}{1}, too.
  This observable gives the maximum value of \chisqd=98.5/5, further
  problems in the determination of \as\ using \momn{\bw}{1} show up in Subsect.~\ref{fitJO}.} 

The fitted values of \asmz\ increase steeply with the order $n$ of the moment used,
for \momn{(1-T)}{n}, \momn{\cp}{n} and \momn{\bt}{n}.  
This effect is less pronounced and systematic
for \momn{\bw}{n}, \momn{(\ytwothree)}{n} and \momn{\mh}{n}. 
In Fig.~\ref{baplot}
we show  the ratio $K={\cal B}_n/{\cal A}_n$ of NLO
and LO coefficients for the six observables used in our fits
to investigate the origin of this behaviour.  
Steeply increasing values of \asmz\ with moment order $n$ 
for \momn{(1-T)}{n}, \momn{\cp}{n} and \momn{\bt}{n}
and
increasing values of $K$ with $n$ 
are clearly correlated. There is also a correlation with the rather large scale 
uncertainties in the respective fits.
The other
observables \momn{\bw}{n}, \momn{(\ytwothree)}{n} and \momn{\mh}{n} 
have more stable results for \asmz\ and
correspondingly fairly 
constant values of $K$.
The reason that the fit of \momn{\mh}{1} does not
converge is the large and negative value of $K$.

\subsection{Combined fits of \JadeBF\ and  \OpalBF\ data}
\label{fitJO}
  For the most significant results we supplement the \Jade\ data
  with the analogous \Opal\ data~\cite{OPALPR404}, covering the energy range of 91 to 209~GeV.

  The \Jade\ and \Opal\ detectors are very similar, both in 
  construction and in the values of many detector parameters.
  The combined use of the \Jade\ and \Opal\ data can therefore be
  expected to lead to consistent measurements, with small 
  systematic differences. Our analysis
  procedure is therefore constructed to be similar to the
  one used in the \Opal\ analysis~\cite{OPALPR404}, in particular in the
  estimate of the systematic errors.

\begin{figure}
     \includegraphics[width=0.48\textwidth]{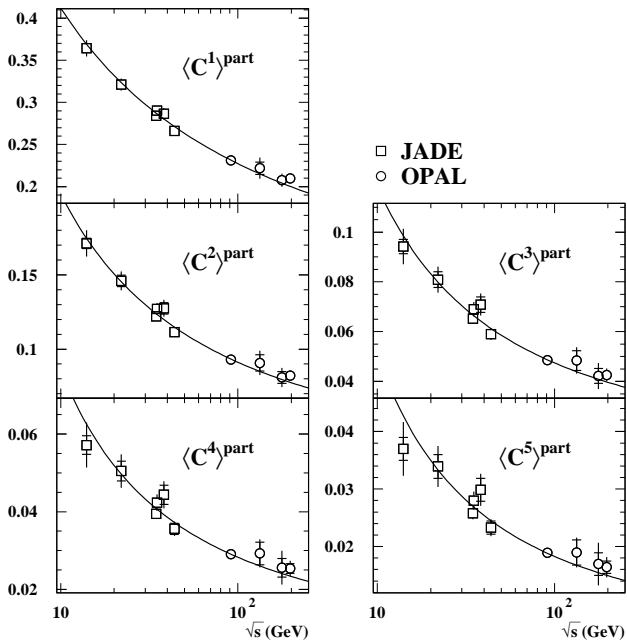}
   \caption{Fits of the NLO predictions to \Jade\ and  \Opal\ measurements of moments of 
            \cp\ at parton level.
  The {\it solid lines} show the \rs\ evolution of the \nlo\ prediction based on the fitted value 
  of \asmz.
  The inner error bars--where visible--show the statistical errors used in the fit,
  the outer bars show the total errors. Where no error bar is visible,
  the total error is smaller than the point size
}
\label{JOIndiCP} 
\end{figure}

\begin{figure}
     \includegraphics[width=0.48\textwidth]{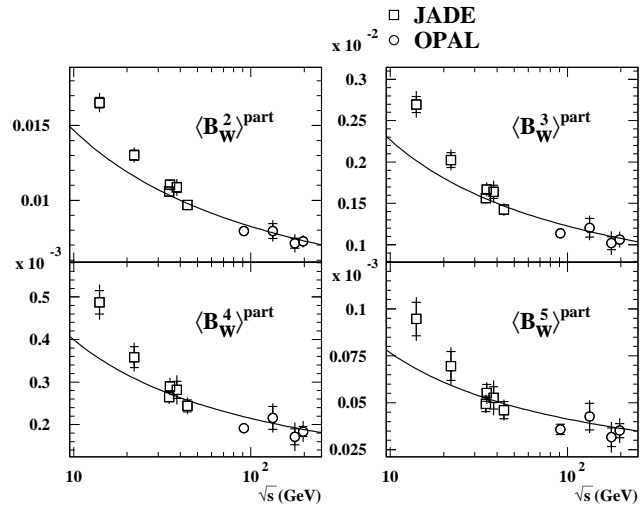}
   \caption{Fits of the NLO predictions to \Jade\ and  \Opal\ measurements of moments of \bw\ at parton level.
  The {\it solid lines} show the \rs\ evolution of the \nlo\ prediction based on the fitted value 
  of \asmz.
  The inner error bars--where visible--show the statistical errors used in the fit,
  the outer bars show the total errors. Where no error bar is visible,
  the total error is smaller than the point size. Problems of the \nlo\ prediction
  at low \rs\ are discussed in the text
}
\label{JOIndiBW} 
\end{figure}

  The central values and statistical errors of the combined fits are found employing the \chisq\ 
calculation~(\ref{simplechi2})
  as above.\footnote{For this reason systematic differences between the two experiments contribute
  to the sometimes high \chisq\ values; in Figs.~\ref{JOIndiCP} and \ref{JOIndiBW} 
  the experimental 
  uncertainties are indicated separately.}
  However, the systematic uncertainties in this case cannot be found by simple repetitions of the fits,
  as the \Jade\ and \Opal\ systematic variations are not identical.

  The systematic uncertainties are correlated between different energy points, and
  including general correlations, the   
  \chisq\ calculation shown in~(\ref{simplechi2}) has to be generalised 
  to \cite{pdg06}
    \begin{eqnarray}
      \label{chi2corrIndiJO}
      \chi^{2} = \sum_{i,j} &&(\momn{y}{n}_i^{\rm part}-\momn{y}{n}^{\rm part,theo}_i)\cdot\nonumber\\ 
               & & V^{-1}_{ij} \cdot(\momn{y}{n}_j^{\rm part}-\momn{y}{n}^{\rm part,theo}_j) \; ,
    \end{eqnarray}
  where the $V_{ij}$ are the covariances
  of the $n$-th moment at the energy points $i$ and $j$. They have the form
 $V_{ij}=S_{ij}+E_{ij}$, with statistical covariances $S_{ij}$ and 
  experimental systematic covariances $E_{ij}$. The matrix $S_{ij}$ is diagonal,
              $S_{ii} = \sigma_{{\rm stat.},\, i}^2\,$,     
  as data of different energy points are independent.
  The experimental systematic covariances $E_{ij}$ are only partly known:      
    \begin{itemize}
      \item The diagonal entries are given by
            \[
              E_{ii}= \sigma_{{\rm exp.},\, i}^2\,,
            \]
            denoting by $\sigma_{\rm exp.,i}$ the experimental uncertainty at energy point
            $i$.       
      \item The non diagonal entries can only follow from plausible assumptions.
            We employ the {\it minimum overlap 
            assumption}\footnote{Fitting the low energy \Jade\ points (14, 22 GeV) 
            this assumption results
            \cite{CHPphd} in a more accurate and more conservative error estimation
            than the {\it full overlap assumption} 
              $E_{ij} = \mathrm{Max}\{\sigma_{{\rm exp.},\,i}^2\,,\;\sigma_{{\rm exp.},\,j}^2\}$
              employed in \cite{OPALPR299}.}     
            \begin{equation}
              E_{ij} = \mathrm{Min}\{\sigma_{{\rm exp.},\,i}^2\,,\;\sigma_{{\rm exp.},\,j}^2\}\,, 
              \label{indiJOmin}
            \end{equation}
    \end{itemize}
  
  The total errors are found by fits employing the \chisq\ calculation (\ref{chi2corrIndiJO}).
  We 
  use the relative experimental uncertainties
  to determine the experimental uncertainties of
  the central values from the fits without correlations.

\begin{figure}[h]
  \begin{center}
  \hspace{-1.5cm}\includegraphics[width=0.48\textwidth]{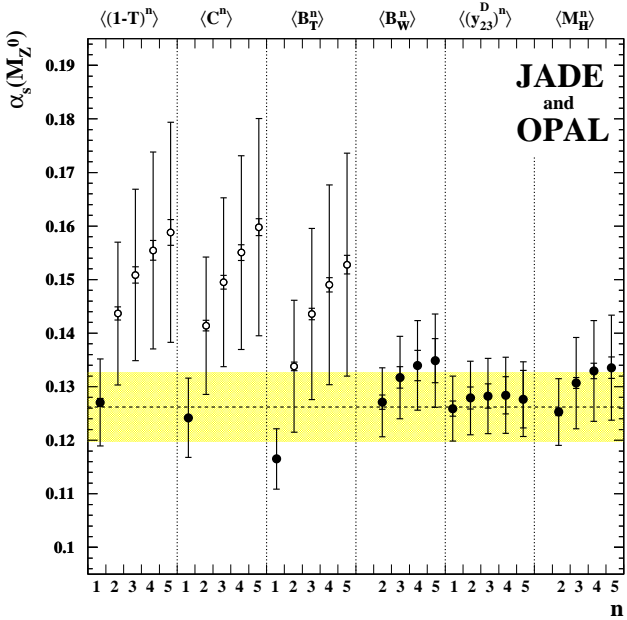}
  \end{center}
  \caption
{Measurements of \asmz\  using fits to moments of six event
  shape variables at \Petra\ and \Lep\ energies.  
  The inner error bars--where visible--show the statistical errors,
  the outer bars show the total errors.
  The experimental systematic uncertainties
  are estimated by the minimum overlap assumption. 
The {\it dotted line} indicates
  the weighted average described in the text, the {\it shaded band} shows its error.
  Only the measurements
  indicated by {\it solid symbols} are used for this purpose}\label{fit_plot_JO}
\label{fit_plotJO}
 \end{figure}

  Figures~\ref{JOIndiCP} and \ref{JOIndiBW} show the comparison of data points and 
  predictions for the moments of the C-parameter and the wide jet broadening \bw.
  The 
  predictions for \momn{\cp}{n} are seen to be
  in good agreement with the data and significantly confirm the QCD prediction of the energy dependence of $\as(\rs)$,
  also known as asymptotic freedom.
    The prediction slightly overshoots the higher moments of \thr, 
    \cp\ and \bt\ at 14\,GeV, and undershoots the moments
    of \bw, \mh,
    and sometimes \ytwothree.
 At low \rs\ the \nlo\ predictions of the
  \bw, \ytwothree\ and \mh\ distributions are (unphysically) negative 
  in a large range of the two jet region 
  \cite{pedrophd}. Therefore the \nlo\ prediction for the moments is likely to be incomplete
  and too low to provide a satisfactory description of the data at low c.m. 
  energies. In the case of \momn{\bw}{1} the $\as^2$ coefficient is even negative, and we 
  do not show or use this fit.
  The results are listed in Table~\ref{JOdpgminovl} and shown in 
  Fig.~\ref{fit_plot_JO}.  

\subsection{Combination of \boldmath{\as} determinations}
\label{ascombs}
To make full use of the data, we combine the measurements of \asmz\ from the
various moments and event shapes and determine a single value.
An extensive study was done by the LEP QCD working group on this 
problem~\cite{MinovlConf,MAF,OPALPR404,STKrev,ALEPH}, and their procedure is adopted here.

A weighted mean of the \asmz\ measurements is calculated 
which minimizes the $\chi^{2}$ formed from the
measurements and the combined value.  
This mean value, \asmz, is given by
\be 
  \asmz=\sum w_{i} \, \asi \;\;\;\; \mathrm{with}\;\;\;\;
  w_{i}=\frac{\sum_{j}(V^{\prime~-1})_{ij}}{\sum_{jk}(V^{\prime~-1})_{jk}},
\ee
where the measured values of \asmz\ 
are denoted \asi, their covariance matrix $V^{\prime}$, 
and the individual results are counted by $i$, $j$ and $k$.  
The presence
of highly correlated and dominant systematic errors
makes
a reliable estimate of $V^{\prime}$ difficult.  
Undesirable features (such as negative\linebreak weights)
can be caused by small
uncertainties in the estimation of these correlations.  
Therefore only
experimental systematic errors--assumed to
be partially correlated by minimum overlap as 
$V^{\prime}_{ij}=\min(\sigma^2_{{\rm exp},i},\sigma^2_{{\rm exp},j})$--and 
statistical correlations are taken to contribute
to the off-diagonal elements of the covariance matrix.
The statistical correlations 
are determined using MC simulations at the
parton level.\footnote{The result is identical if the correlations are determined
using \py, \hw\ or \ar\ at 14.0...43.8 GeV, or determined at hadron level instead of parton level. 
The correlation values are cited
in~\cite{CHPphd}; at 14~GeV and parton level they vary between 29\% and 99\% and are larger than 50\% mostly.} The 
diagonal elements 
are calculated from
all error
contributions--statistical, experimental, hadronisation and theory uncertainties.
Using the weights derived from the covariance matrix $V^{\prime}$ the
theory uncertainties are computed by analogously combining the
\asmz\ values
from setting $\xmu=2.0$ or $\xmu=0.5$, 
and the hadronisation uncertainties by combining the results obtained with the 
alternative hadronisation models.

\begin{figure}
   \begin{center}
   \includegraphics[width=0.45\textwidth]{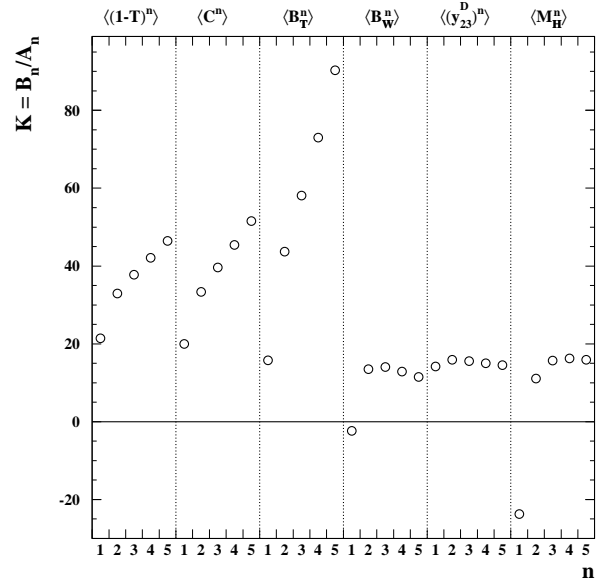}
   \end{center}
   \caption
   {The ratio $K=\mathcal{B}_n/\mathcal{A}_n$
     of \nlo\ and \lo\ coefficients for the first five moments of the six event shape variables
     used in the determination of \as, see also \cite{OPALPR404}
   }
\label{baplot}
\end{figure}

To select observables with an
apparently converging perturbative prediction, we consider \cite{OPALPR404} only those results for which the NLO term in
equation~(\ref{eq_qcdmom}) is less than half the corresponding LO term (i.e.\ 
$|K\as/2\pi|<0.5$ or $|K|<25$), namely
\momone{\thr}, \momone{\cp}, \momone{\bt}, \momn{\bw}{n} and
\momn{(\ytwothree)}{n}, $n=1,\ldots,5$; and \momn{\mh}{n},
$n=2,\ldots,5$. These are results from 17 observables in total;
or 16 observables from \Jade\ and \Opal, exclu\-ding \momn{\bw}{1}.  
  The $K$ values are
shown in Fig.~\ref{baplot}.  

Using only \Jade\ data, the result of the combination is
 \resJlines
and is shown in Fig.~\ref{fit_plot}. 
Combining \Jade\ and \Opal\ measurements, the result is
  \resJOlines
and is shown in Fig.~\ref{fit_plotJO}.
Both values are above, but still consistent with the world average of
$\asmz=0.1189\pm0.0010$~\cite{bethke06}.  It has been observed previously
in comparisons of event shape distributions  with NLO~\cite{OPALPR075} or\linebreak
NNLO~\cite{asNNLO}
QCD predictions with $\xmu=1$ that fitted values of \asmz\ tend to
be large compared to the world average.

To enable comparison with earlier and more specific analyses \cite{JADE-paper} we
combine the \Jade\ fit results from only the first\footnote{Because of the problems with the 
\nlo\ description of \momn{\mh}{1}$^{\rm part}$, \momn{\mh}{2} is often regarded as the first moment of \mh.} 
moments \momone{\thr}, \momone{\cp}, \momone{\bt}, \momone{\bw}, \momone{\ytwothree} and \momn{\mh}{2}. 
This yields a value of
\begin{eqnarray*}
  \asmz&=&0.1243\pm0.0001\stat\pm0.0009\expt\nonumber\\
       & &\pm0.0010\had\pm0.0070\theo \,.
\end{eqnarray*}
The slightly smaller error in this determination of \as\ reflects the fact that the lower
order moments are less sensitive to the multijet region of the 
event shape distributions. This leads to a smaller statistical and theoretical uncertainty.
In all three measurements the scale uncertainty is dominant.

\section{Summary}
\label{summary}
In this paper we present measurements of moments of event shape distributions 
at centre-of-mass
energies between 14 and 44~GeV using data of the \Jade\ experiment. The
predictions of the \py, \hw\ and \ar\ MC models tuned by \Opal\ 
to LEP~1 data are found to be in reasonable agreement with the measured
moments.

From fits of \oaa\ predictions to selected event shape moments
corrected for experimental and hadronisation effects
we have
determined the strong coupling to be\linebreak
  \resJtot\ using only \Jade\ data, and
  \resJOtot\ using combined \Jade\ and \Opal\ data.
Fits to moments of \mh, \bw\ and \ytwothree\ return large values of \chisqd;
the higher moments, in particular of the \thr, \cp\ and
\bt\ event shape variables, lead to systematically enlarged values of 
\as.
Results where \as\ is steeply rising with moment order are strongly correlated with the relative
size of the \asbsq\ coefficient and thus are most likely affected by deficiencies of the \nlo\ prediction.

The \Jade\ experiment assesses an interesting energy range for the perturbative
analysis since the energy evolution of the strong coupling is more pronounced at
low energies. 

{\small
\section*{Acknowledgements}
\par
This research was supported by the DFG cluster of excellence `Origin
and Structure of the Universe'.
}

 \bibliographystyle{iopart}
 \bibliography{papers}

\begin{table*}
\caption{Moments of the $1-T$, $C$, \bt, \bw, \ytwothree\ and \mh\ distributions measured by
\Jade\ at 14.0, 22.0 and 34.6~GeV.
The first uncertainty is statistical, while the second is systematic}\label{tabmomA}
\begin{center}
\begin{tabular}{ c  r @{(} r @{.} l @{ $\pm$ } r @{.} l @{ $\pm$ } r @{.} l @{)} l @{} l    
                    r @{(} r @{.} l @{ $\pm$ } r @{.} l @{ $\pm$ } r @{.} l @{)} l @{} l    
                    r @{(} r @{.} l @{ $\pm$ } r @{.} l @{ $\pm$ } r @{.} l @{)} l @{} l    
                    r @{(} r @{.} l @{ $\pm$ } r @{.} l @{ $\pm$ } r @{.} l @{)} l @{} l    
                    r @{(} r @{.} l @{ $\pm$ } r @{.} l @{ $\pm$ } r @{.} l @{)} l @{} l   }
 \hline\noalign{\smallskip} \multicolumn{1}{ c }{$n$} & \multicolumn{9}{ c }{\momn{(\thr)}{n} at 14.0 GeV} & \multicolumn{9}{ c }{\momn{(\thr)}{n} at 22.0 GeV} & \multicolumn{9}{ c }{\momn{(\thr)}{n} at 34.6 GeV} \\
 \noalign{\smallskip}\hline\noalign{\smallskip} 
 $1$  & & 1&405 & 0&022 & 0&050 & $\cdot 10^{-1}$ &  & & 1&123 & 0&021 & 0&028 & $\cdot 10^{-1}$ &  & & 8&99 & 0&07 & 0&13 & $\cdot 10^{-2}$ & \\
 $2$  & & 2&38 & 0&08 & 0&17 & $\cdot 10^{-2}$ &  & & 1&700 & 0&068 & 0&086 & $\cdot 10^{-2}$ &  & & 1&192 & 0&020 & 0&024 & $\cdot 10^{-2}$ & \\
 $3$  & & 4&68 & 0&28 & 0&54 & $\cdot 10^{-3}$ &  & & 3&31 & 0&21 & 0&26 & $\cdot 10^{-3}$ &  & & 2&151 & 0&057 & 0&052 & $\cdot 10^{-3}$ & \\
 $4$  & & 1&04 & 0&09 & 0&17 & $\cdot 10^{-3}$ &  & & 7&79 & 0&69 & 0&84 & $\cdot 10^{-4}$ &  & & 4&77 & 0&17 & 0&16 & $\cdot 10^{-4}$ & \\
 $5$  & & 2&55 & 0&33 & 0&56 & $\cdot 10^{-4}$ &  & & 2&08 & 0&24 & 0&29 & $\cdot 10^{-4}$ &  & & 1&202 & 0&056 & 0&061 & $\cdot 10^{-4}$ &  \\ \noalign{\smallskip}\hline\hline\noalign{\smallskip} \multicolumn{1}{ c }{$n$} & \multicolumn{9}{ c }{\momn{\cp}{n} at 14.0 GeV} & \multicolumn{9}{ c }{\momn{\cp}{n} at 22.0 GeV} & \multicolumn{9}{ c }{\momn{\cp}{n} at 34.6 GeV} \\
 \noalign{\smallskip}\hline\noalign{\smallskip} 
 $1$  & & 5&22 & 0&05 & 0&13 & $\cdot 10^{-1}$ &  & & 4&280 & 0&057 & 0&077 & $\cdot 10^{-1}$ &  & & 3&512 & 0&020 & 0&043 & $\cdot 10^{-1}$ & \\
 $2$  & & 3&00 & 0&06 & 0&14 & $\cdot 10^{-1}$ &  & & 2&152 & 0&056 & 0&075 & $\cdot 10^{-1}$ &  & & 1&561 & 0&018 & 0&030 & $\cdot 10^{-1}$ & \\
 $3$  & & 1&85 & 0&06 & 0&13 & $\cdot 10^{-1}$ &  & & 1&235 & 0&047 & 0&064 & $\cdot 10^{-1}$ &  & & 8&31 & 0&14 & 0&19 & $\cdot 10^{-2}$ & \\
 $4$  & & 1&22 & 0&05 & 0&11 & $\cdot 10^{-1}$ &  & & 7&86 & 0&39 & 0&54 & $\cdot 10^{-2}$ &  & & 5&04 & 0&11 & 0&13 & $\cdot 10^{-2}$ & \\
 $5$  & & 8&39 & 0&45 & 0&96 & $\cdot 10^{-2}$ &  & & 5&41 & 0&33 & 0&46 & $\cdot 10^{-2}$ &  & & 3&335 & 0&088 & 0&099 & $\cdot 10^{-2}$ &  \\ \noalign{\smallskip}\hline\hline\noalign{\smallskip} \multicolumn{1}{ c }{$n$} & \multicolumn{9}{ c }{\momn{\bt}{n} at 14.0 GeV} & \multicolumn{9}{ c }{\momn{\bt}{n} at 22.0 GeV} & \multicolumn{9}{ c }{\momn{\bt}{n} at 34.6 GeV} \\
 \noalign{\smallskip}\hline\noalign{\smallskip} 
 $1$  & & 1&918 & 0&017 & 0&038 & $\cdot 10^{-1}$ &  & & 1&627 & 0&018 & 0&021 & $\cdot 10^{-1}$ &  & & 1&372 & 0&006 & 0&011 & $\cdot 10^{-1}$ & \\
 $2$  & & 3&94 & 0&07 & 0&16 & $\cdot 10^{-2}$ &  & & 2&963 & 0&067 & 0&077 & $\cdot 10^{-2}$ &  & & 2&202 & 0&021 & 0&033 & $\cdot 10^{-2}$ & \\
 $3$  & & 8&61 & 0&26 & 0&54 & $\cdot 10^{-3}$ &  & & 6&01 & 0&21 & 0&24 & $\cdot 10^{-3}$ &  & & 4&082 & 0&062 & 0&083 & $\cdot 10^{-3}$ & \\
 $4$  & & 1&99 & 0&09 & 0&17 & $\cdot 10^{-3}$ &  & & 1&344 & 0&065 & 0&072 & $\cdot 10^{-3}$ &  & & 8&56 & 0&18 & 0&22 & $\cdot 10^{-4}$ & \\
 $5$  & & 4&84 & 0&28 & 0&55 & $\cdot 10^{-4}$ &  & & 3&26 & 0&20 & 0&22 & $\cdot 10^{-4}$ &  & & 1&978 & 0&053 & 0&062 & $\cdot 10^{-4}$ &  \\ \noalign{\smallskip}\hline\hline\noalign{\smallskip} \multicolumn{1}{ c }{$n$} & \multicolumn{9}{ c }{\momn{\bw}{n} at 14.0 GeV} & \multicolumn{9}{ c }{\momn{\bw}{n} at 22.0 GeV} & \multicolumn{9}{ c }{\momn{\bw}{n} at 34.6 GeV} \\
 \noalign{\smallskip}\hline\noalign{\smallskip} 
 $1$  & & 1&166 & 0&011 & 0&019 & $\cdot 10^{-1}$ &  & & 1&000 & 0&012 & 0&014 & $\cdot 10^{-1}$ &  & & 8&720 & 0&047 & 0&087 & $\cdot 10^{-2}$ & \\
 $2$  & & 1&482 & 0&031 & 0&048 & $\cdot 10^{-2}$ &  & & 1&151 & 0&030 & 0&033 & $\cdot 10^{-2}$ &  & & 9&42 & 0&11 & 0&20 & $\cdot 10^{-3}$ & \\
 $3$  & & 2&045 & 0&070 & 0&098 & $\cdot 10^{-3}$ &  & & 1&525 & 0&065 & 0&068 & $\cdot 10^{-3}$ &  & & 1&238 & 0&023 & 0&042 & $\cdot 10^{-3}$ & \\
 $4$  & & 3&04 & 0&15 & 0&19 & $\cdot 10^{-4}$ &  & & 2&30 & 0&14 & 0&15 & $\cdot 10^{-4}$ &  & & 1&897 & 0&050 & 0&085 & $\cdot 10^{-4}$ & \\
 $5$  & & 4&81 & 0&33 & 0&38 & $\cdot 10^{-5}$ &  & & 3&84 & 0&32 & 0&36 & $\cdot 10^{-5}$ &  & & 3&25 & 0&11 & 0&18 & $\cdot 10^{-5}$ &  \\ \noalign{\smallskip}\hline\hline\noalign{\smallskip} \multicolumn{1}{ c }{$n$} & \multicolumn{9}{ c }{\momn{(\ytwothree)}{n} at 14.0 GeV} & \multicolumn{9}{ c }{\momn{(\ytwothree)}{n} at 22.0 GeV} & \multicolumn{9}{ c }{\momn{(\ytwothree)}{n} at 34.6 GeV} \\
 \noalign{\smallskip}\hline\noalign{\smallskip} 
 $1$  & & 3&54 & 0&12 & 0&19 & $\cdot 10^{-2}$ &  & & 2&89 & 0&12 & 0&10 & $\cdot 10^{-2}$ &  & & 2&408 & 0&042 & 0&041 & $\cdot 10^{-2}$ & \\
 $2$  & & 2&55 & 0&22 & 0&29 & $\cdot 10^{-3}$ &  & & 2&55 & 0&23 & 0&18 & $\cdot 10^{-3}$ &  & & 2&173 & 0&081 & 0&049 & $\cdot 10^{-3}$ & \\
 $3$  & & 3&01 & 0&48 & 0&48 & $\cdot 10^{-4}$ &  & & 3&93 & 0&55 & 0&37 & $\cdot 10^{-4}$ &  & & 3&35 & 0&19 & 0&13 & $\cdot 10^{-4}$ & \\
 $4$  & & 4&7 & 1&2 & 1&0 & $\cdot 10^{-5}$ &  & & 7&6 & 1&4 & 0&8 & $\cdot 10^{-5}$ &  & & 6&42 & 0&48 & 0&38 & $\cdot 10^{-5}$ & \\
 $5$  & & 8&8 & 3&0 & 2&5 & $\cdot 10^{-6}$ &  & & 1&66 & 0&36 & 0&21 & $\cdot 10^{-5}$ &  & & 1&38 & 0&13 & 0&11 & $\cdot 10^{-5}$ &  \\ \noalign{\smallskip}\hline\hline\noalign{\smallskip} \multicolumn{1}{ c }{$n$} & \multicolumn{9}{ c }{\momn{\mh}{n} at 14.0 GeV} & \multicolumn{9}{ c }{\momn{\mh}{n} at 22.0 GeV} & \multicolumn{9}{ c }{\momn{\mh}{n} at 34.6 GeV} \\
 \noalign{\smallskip}\hline\noalign{\smallskip} 
 $1$  & & 3&207 & 0&024 & 0&049 & $\cdot 10^{-1}$ &  & & 2&832 & 0&026 & 0&036 & $\cdot 10^{-1}$ &  & & 2&522 & 0&010 & 0&024 & $\cdot 10^{-1}$ & \\
 $2$  & & 1&074 & 0&017 & 0&033 & $\cdot 10^{-1}$ &  & & 8&55 & 0&16 & 0&21 & $\cdot 10^{-2}$ &  & & 6&979 & 0&057 & 0&095 & $\cdot 10^{-2}$ & \\
 $3$  & & 3&74 & 0&09 & 0&17 & $\cdot 10^{-2}$ &  & & 2&74 & 0&08 & 0&10 & $\cdot 10^{-2}$ &  & & 2&114 & 0&028 & 0&030 & $\cdot 10^{-2}$ & \\
 $4$  & & 1&348 & 0&048 & 0&085 & $\cdot 10^{-2}$ &  & & 9&36 & 0&40 & 0&49 & $\cdot 10^{-3}$ &  & & 6&98 & 0&13 & 0&10 & $\cdot 10^{-3}$ & \\
 $5$  & & 5&02 & 0&24 & 0&41 & $\cdot 10^{-3}$ &  & & 3&38 & 0&20 & 0&23 & $\cdot 10^{-3}$ &  & & 2&485 & 0&062 & 0&051 & $\cdot 10^{-3}$ &  \\ \hline\noalign{\smallskip}
\end{tabular}
\end{center}
\end{table*}

\begin{table*}
\caption{Moments of the $1-T$, $C$, \bt, \bw, \ytwothree\ and \mh\ distributions measured by
\Jade\ at 35.0, 38.3 and 43.8~GeV.
The first uncertainty is statistical, while the second is systematic}\label{tabmomB}
\begin{center}
\begin{tabular}{ c  r @{(} r @{.} l @{ $\pm$ } r @{.} l @{ $\pm$ } r @{.} l @{)} l @{} l    
                    r @{(} r @{.} l @{ $\pm$ } r @{.} l @{ $\pm$ } r @{.} l @{)} l @{} l    
                    r @{(} r @{.} l @{ $\pm$ } r @{.} l @{ $\pm$ } r @{.} l @{)} l @{} l    
                    r @{(} r @{.} l @{ $\pm$ } r @{.} l @{ $\pm$ } r @{.} l @{)} l @{} l    
                    r @{(} r @{.} l @{ $\pm$ } r @{.} l @{ $\pm$ } r @{.} l @{)} l @{} l   }
 \hline\noalign{\smallskip} \multicolumn{1}{ c }{$n$} & \multicolumn{9}{ c }{\momn{(\thr)}{n} at 35.0 GeV} & \multicolumn{9}{ c }{\momn{(\thr)}{n} at 38.3 GeV} & \multicolumn{9}{ c }{\momn{(\thr)}{n} at 43.8 GeV} \\
 \noalign{\smallskip}\hline\noalign{\smallskip} 
 $1$  & & 9&22 & 0&07 & 0&18 & $\cdot 10^{-2}$ &  & & 9&06 & 0&19 & 0&22 & $\cdot 10^{-2}$ &  & & 8&07 & 0&12 & 0&10 & $\cdot 10^{-2}$ & \\
 $2$  & & 1&260 & 0&019 & 0&045 & $\cdot 10^{-2}$ &  & & 1&266 & 0&056 & 0&061 & $\cdot 10^{-2}$ &  & & 1&032 & 0&032 & 0&024 & $\cdot 10^{-2}$ & \\
 $3$  & & 2&34 & 0&06 & 0&11 & $\cdot 10^{-3}$ &  & & 2&41 & 0&16 & 0&17 & $\cdot 10^{-3}$ &  & & 1&867 & 0&093 & 0&069 & $\cdot 10^{-3}$ & \\
 $4$  & & 5&36 & 0&17 & 0&29 & $\cdot 10^{-4}$ &  & & 5&56 & 0&50 & 0&53 & $\cdot 10^{-4}$ &  & & 4&22 & 0&29 & 0&23 & $\cdot 10^{-4}$ & \\
 $5$  & & 1&394 & 0&058 & 0&084 & $\cdot 10^{-4}$ &  & & 1&44 & 0&16 & 0&17 & $\cdot 10^{-4}$ &  & & 1&099 & 0&098 & 0&084 & $\cdot 10^{-4}$ &  \\ \noalign{\smallskip}\hline\hline\noalign{\smallskip} \multicolumn{1}{ c }{$n$} & \multicolumn{9}{ c }{\momn{\cp}{n} at 35.0 GeV} & \multicolumn{9}{ c }{\momn{\cp}{n} at 38.3 GeV} & \multicolumn{9}{ c }{\momn{\cp}{n} at 43.8 GeV} \\
 \noalign{\smallskip}\hline\noalign{\smallskip} 
 $1$  & & 3&582 & 0&019 & 0&057 & $\cdot 10^{-1}$ &  & & 3&486 & 0&056 & 0&065 & $\cdot 10^{-1}$ &  & & 3&178 & 0&034 & 0&032 & $\cdot 10^{-1}$ & \\
 $2$  & & 1&620 & 0&017 & 0&047 & $\cdot 10^{-1}$ &  & & 1&587 & 0&048 & 0&052 & $\cdot 10^{-1}$ &  & & 1&347 & 0&028 & 0&025 & $\cdot 10^{-1}$ & \\
 $3$  & & 8&76 & 0&13 & 0&34 & $\cdot 10^{-2}$ &  & & 8&76 & 0&38 & 0&39 & $\cdot 10^{-2}$ &  & & 7&05 & 0&22 & 0&18 & $\cdot 10^{-2}$ & \\
 $4$  & & 5&38 & 0&10 & 0&25 & $\cdot 10^{-2}$ &  & & 5&49 & 0&30 & 0&30 & $\cdot 10^{-2}$ &  & & 4&25 & 0&17 & 0&14 & $\cdot 10^{-2}$ & \\
 $5$  & & 3&60 & 0&08 & 0&19 & $\cdot 10^{-2}$ &  & & 3&74 & 0&25 & 0&24 & $\cdot 10^{-2}$ &  & & 2&81 & 0&14 & 0&11 & $\cdot 10^{-2}$ &  \\ \noalign{\smallskip}\hline\hline\noalign{\smallskip} \multicolumn{1}{ c }{$n$} & \multicolumn{9}{ c }{\momn{\bt}{n} at 35.0 GeV} & \multicolumn{9}{ c }{\momn{\bt}{n} at 38.3 GeV} & \multicolumn{9}{ c }{\momn{\bt}{n} at 43.8 GeV} \\
 \noalign{\smallskip}\hline\noalign{\smallskip} 
 $1$  & & 1&395 & 0&006 & 0&016 & $\cdot 10^{-1}$ &  & & 1&364 & 0&018 & 0&019 & $\cdot 10^{-1}$ &  & & 1&260 & 0&011 & 0&010 & $\cdot 10^{-1}$ & \\
 $2$  & & 2&277 & 0&020 & 0&054 & $\cdot 10^{-2}$ &  & & 2&229 & 0&059 & 0&064 & $\cdot 10^{-2}$ &  & & 1&920 & 0&035 & 0&031 & $\cdot 10^{-2}$ & \\
 $3$  & & 4&30 & 0&06 & 0&15 & $\cdot 10^{-3}$ &  & & 4&27 & 0&17 & 0&18 & $\cdot 10^{-3}$ &  & & 3&480 & 0&098 & 0&087 & $\cdot 10^{-3}$ & \\
 $4$  & & 9&20 & 0&17 & 0&42 & $\cdot 10^{-4}$ &  & & 9&30 & 0&51 & 0&54 & $\cdot 10^{-4}$ &  & & 7&26 & 0&28 & 0&26 & $\cdot 10^{-4}$ & \\
 $5$  & & 2&17 & 0&05 & 0&12 & $\cdot 10^{-4}$ &  & & 2&23 & 0&15 & 0&16 & $\cdot 10^{-4}$ &  & & 1&681 & 0&084 & 0&080 & $\cdot 10^{-4}$ &  \\ \noalign{\smallskip}\hline\hline\noalign{\smallskip} \multicolumn{1}{ c }{$n$} & \multicolumn{9}{ c }{\momn{\bw}{n} at 35.0 GeV} & \multicolumn{9}{ c }{\momn{\bw}{n} at 38.3 GeV} & \multicolumn{9}{ c }{\momn{\bw}{n} at 43.8 GeV} \\
 \noalign{\smallskip}\hline\noalign{\smallskip} 
 $1$  & & 8&90 & 0&04 & 0&13 & $\cdot 10^{-2}$ &  & & 8&76 & 0&13 & 0&17 & $\cdot 10^{-2}$ &  & & 8&185 & 0&082 & 0&097 & $\cdot 10^{-2}$ & \\
 $2$  & & 9&83 & 0&10 & 0&31 & $\cdot 10^{-3}$ &  & & 9&72 & 0&29 & 0&40 & $\cdot 10^{-3}$ &  & & 8&73 & 0&19 & 0&22 & $\cdot 10^{-3}$ & \\
 $3$  & & 1&323 & 0&022 & 0&061 & $\cdot 10^{-3}$ &  & & 1&319 & 0&062 & 0&087 & $\cdot 10^{-3}$ &  & & 1&171 & 0&040 & 0&046 & $\cdot 10^{-3}$ & \\
 $4$  & & 2&08 & 0&05 & 0&12 & $\cdot 10^{-4}$ &  & & 2&07 & 0&13 & 0&20 & $\cdot 10^{-4}$ &  & & 1&859 & 0&089 & 0&100 & $\cdot 10^{-4}$ & \\
 $5$  & & 3&65 & 0&11 & 0&24 & $\cdot 10^{-5}$ &  & & 3&60 & 0&30 & 0&50 & $\cdot 10^{-5}$ &  & & 3&31 & 0&21 & 0&23 & $\cdot 10^{-5}$ &  \\ \noalign{\smallskip}\hline\hline\noalign{\smallskip} \multicolumn{1}{ c }{$n$} & \multicolumn{9}{ c }{\momn{(\ytwothree)}{n} at 35.0 GeV} & \multicolumn{9}{ c }{\momn{(\ytwothree)}{n} at 38.3 GeV} & \multicolumn{9}{ c }{\momn{(\ytwothree)}{n} at 43.8 GeV} \\
 \noalign{\smallskip}\hline\noalign{\smallskip} 
 $1$  & & 2&551 & 0&040 & 0&058 & $\cdot 10^{-2}$ &  & & 2&66 & 0&12 & 0&16 & $\cdot 10^{-2}$ &  & & 2&269 & 0&071 & 0&068 & $\cdot 10^{-2}$ & \\
 $2$  & & 2&395 & 0&080 & 0&071 & $\cdot 10^{-3}$ &  & & 2&62 & 0&23 & 0&28 & $\cdot 10^{-3}$ &  & & 2&10 & 0&14 & 0&11 & $\cdot 10^{-3}$ & \\
 $3$  & & 3&77 & 0&19 & 0&23 & $\cdot 10^{-4}$ &  & & 4&04 & 0&52 & 0&61 & $\cdot 10^{-4}$ &  & & 3&24 & 0&32 & 0&28 & $\cdot 10^{-4}$ & \\
 $4$  & & 7&31 & 0&48 & 0&71 & $\cdot 10^{-5}$ &  & & 7&5 & 1&3 & 1&6 & $\cdot 10^{-5}$ &  & & 6&19 & 0&82 & 0&77 & $\cdot 10^{-5}$ & \\
 $5$  & & 1&58 & 0&13 & 0&20 & $\cdot 10^{-5}$ &  & & 1&52 & 0&34 & 0&44 & $\cdot 10^{-5}$ &  & & 1&33 & 0&22 & 0&20 & $\cdot 10^{-5}$ &  \\ \noalign{\smallskip}\hline\hline\noalign{\smallskip} \multicolumn{1}{ c }{$n$} & \multicolumn{9}{ c }{\momn{\mh}{n} at 35.0 GeV} & \multicolumn{9}{ c }{\momn{\mh}{n} at 38.3 GeV} & \multicolumn{9}{ c }{\momn{\mh}{n} at 43.8 GeV} \\
 \noalign{\smallskip}\hline\noalign{\smallskip} 
 $1$  & & 2&555 & 0&009 & 0&018 & $\cdot 10^{-1}$ &  & & 2&509 & 0&027 & 0&018 & $\cdot 10^{-1}$ &  & & 2&371 & 0&017 & 0&026 & $\cdot 10^{-1}$ & \\
 $2$  & & 7&174 & 0&052 & 0&081 & $\cdot 10^{-2}$ &  & & 7&00 & 0&15 & 0&11 & $\cdot 10^{-2}$ &  & & 6&316 & 0&097 & 0&098 & $\cdot 10^{-2}$ & \\
 $3$  & & 2&209 & 0&026 & 0&035 & $\cdot 10^{-2}$ &  & & 2&154 & 0&074 & 0&064 & $\cdot 10^{-2}$ &  & & 1&884 & 0&047 & 0&034 & $\cdot 10^{-2}$ & \\
 $4$  & & 7&42 & 0&12 & 0&17 & $\cdot 10^{-3}$ &  & & 7&22 & 0&35 & 0&39 & $\cdot 10^{-3}$ &  & & 6&24 & 0&22 & 0&14 & $\cdot 10^{-3}$ & \\
 $5$  & & 2&694 & 0&060 & 0&091 & $\cdot 10^{-3}$ &  & & 2&61 & 0&16 & 0&22 & $\cdot 10^{-3}$ &  & & 2&26 & 0&11 & 0&07 & $\cdot 10^{-3}$ &  \\ \noalign{\smallskip}\hline
\end{tabular}
\end{center}
\end{table*}

\begin{table*}
 \caption{Measurements of \asmz\  from event shape moments
   over the full analysed range of \Petra\ c.m.\ energies, 14--44\,GeV. 
   The hadronisation uncertainty is taken to 
   be the larger of the deviations observed using \Herwig\ and \Ariadne}
 \label{JADE-note-dt}
\begin{center}
\small
\resizebox{0.69\textwidth}{!}{
\begin{tabular}{  l   r@{}l r@{}l r@{}l r@{}l r@{}l r@{}l   r@{}l  }
\hline\noalign{\smallskip} 
&\multicolumn{2}{c}{$\langle (1-T)^1 \rangle$} 
&\multicolumn{2}{c}{$\langle C^1 \rangle$} 
&\multicolumn{2}{c}{\momn{\bt}{1}} 
&\multicolumn{2}{c}{\momn{\bw}{1}} 
&\multicolumn{2}{c}{$\langle (\ytwothree)^1 \rangle$} 
&\multicolumn{2}{c }{} 
 \\ 
 
 \noalign{\smallskip}\hline\noalign{\smallskip} 
{\bf \boldmath \asmz}&$0.1276$&&$0.1241$&&$0.1157$&&$0.1308$&&$0.1346$&&&
 \\ 
Statistical error&$0.0004$&&$0.0003$&&$0.0002$&&$0.0004$&&$0.0009$&&&
 \\ 
Experimental syst.&$0.0013$&&$0.0010$&&$0.0006$&&$0.0014$&&$0.0016$&&&
 \\ 
HERWIG hadr.~corr. &$-0.0017$&&$-0.0017$&&$-0.0003$&    &$-0.0007$&    &$+0.0011$&&&
 \\ 
ARIADNE hadr.~corr.&$+0.0002$&    &$+0.0000$&    &$+0.0009$&&$-0.0042$&&$-0.0051$&&&
 \\ 
Hadronisation error &$ 0.0017$&    &$ 0.0017$&    &$ 0.0009$&    &$ 0.0042$&    &$ 0.0051$&&&
 \\ 
$x_\mu$ variation: \\
\hspace{.3cm}${\xmu=2.0} $&${+0.0084} $&&${+0.0076} $&&${+0.0055} $&&${+0.0097} $&&${+0.0079} $&&&\\
\hspace{.3cm}${\xmu=0.5}$&$ {-0.0068}$&&$ {-0.0061}$&&$ {-0.0043}$&&$ {-0.0005}$&&$ {-0.0059}$&&&
 \\ 
\chisqd&$14.9/5$&&$16.7/5$&&$48.8/5$&&$98.8/5$&&$40.0/5$&&&
 \\ 
\noalign{\smallskip}\hline\hline\noalign{\smallskip} 
&\multicolumn{2}{c}{$\langle (1-T)^2 \rangle$} 
&\multicolumn{2}{c}{$\langle C^2 \rangle$} 
&\multicolumn{2}{c}{\momn{\bt}{2}} 
&\multicolumn{2}{c}{\momn{\bw}{2}} 
&\multicolumn{2}{c}{$\langle (\ytwothree)^2 \rangle$} 
&\multicolumn{2}{c }{\momn{\mh}{2}} 
 \\ 
 
 \noalign{\smallskip}\hline\noalign{\smallskip} 
{\bf \boldmath \asmz}&$0.1447$&&$0.1417$&&$0.1333$&&$0.1327$&&$0.1369$&&$0.1294$&
 \\ 
Statistical error&$0.0008$&&$0.0005$&&$0.0004$&&$0.0006$&&$0.0019$&&$0.0004$&
 \\ 
Experimental syst.&$0.0019$&&$0.0017$&&$0.0011$&&$0.0021$&&$0.0016$&&$0.0011$&
 \\ 
HERWIG hadr.~corr. &$+0.0009$&&$-0.0001$& &$+0.0006$& &$-0.0006$& &$+0.0026$& &$+0.0051$& 
 \\ 
ARIADNE hadr.~corr.&$+0.0009$& &$+0.0007$&&$+0.0011$&&$-0.0048$&&$-0.0043$&&$-0.0024$&
 \\ 
Hadronisation error &$ 0.0009$& &$ 0.0007$& &$ 0.0011$& &$ 0.0048$& &$ 0.0043$& &$0.0051$&
 \\ 
$x_\mu$ variation: \\
\hspace{.3cm}${\xmu=2.0}$&${+0.0141} $&&${+0.0134} $&&${+0.0125} $&&${+0.0074} $&&${+0.0088} $&&${+0.0062} $&\\
\hspace{.3cm}${\xmu=0.5}$&${-0.0113}$&&$ {-0.0109}$&&$ {-0.0103}$&&$ {-0.0055}$&&$ {-0.0067}$&&$ {-0.0043}$&
 \\ 
\chisqd&$13.5/5$&&$16.3/5$&&$33.7/5$&&$64.7/5$&&$13.7/5$&&$92.7/5$&
 \\ 
\noalign{\smallskip}\hline\hline\noalign{\smallskip} 
&\multicolumn{2}{c}{$\langle (1-T)^3 \rangle$} 
&\multicolumn{2}{c}{$\langle C^3 \rangle$} 
&\multicolumn{2}{c}{\momn{\bt}{3}} 
&\multicolumn{2}{c}{\momn{\bw}{3}} 
&\multicolumn{2}{c}{$\langle (\ytwothree)^3 \rangle$} 
&\multicolumn{2}{c }{\momn{\mh}{3}} 
 \\ 
 
 \noalign{\smallskip}\hline\noalign{\smallskip} 
{\bf \boldmath \asmz}&$0.1514$&&$0.1497$&&$0.1434$&&$0.1376$&&$0.1352$&&$0.1364$&
 \\ 
Statistical error&$0.0013$&&$0.0007$&&$0.0007$&&$0.0011$&&$0.0030$&&$0.0007$&
 \\ 
Experimental syst.&$0.0022$&&$0.0021$&&$0.0014$&&$0.0032$&&$0.0027$&&$0.0012$&
 \\ 
HERWIG hadr.~corr. &$+0.0033$&&$+0.0016$&&$+0.0018$&&$-0.0006$& &$+0.0033$& &$+0.0069$& 
 \\ 
ARIADNE hadr.~corr.&$+0.0016$& &$+0.0015$& &$+0.0012$& &$-0.0059$&&$-0.0039$&&$-0.0030$&
 \\ 
Hadronisation error &$ 0.0033$& &$ 0.0016$& &$ 0.0018$& &$ 0.0059$& &$ 0.0039$& &$0.0069$&
 \\ 
$x_\mu$ variation:\\
\hspace{.3cm}${\xmu=2.0} $&${+0.0166} $&&${+0.0164} $&&${+0.0162} $&&${+0.0084} $&&${+0.0084} $&&${+0.0087} $&\\
\hspace{.3cm}${\xmu=0.5}$&$ {-0.0132}$&&$ {-0.0131}$&&$ {-0.0130}$&&$ {-0.0063}$&&$ {-0.0064}$&&$ {-0.0067}$&
 \\ 
\chisqd&$12.1/5$&&$16.5/5$&&$23.8/5$&&$43.8/5$&&$6.0/5$&&$66.9/5$&
 \\ 
\noalign{\smallskip}\hline\hline\noalign{\smallskip} 
&\multicolumn{2}{c}{$\langle (1-T)^4 \rangle$} 
&\multicolumn{2}{c}{$\langle C^4 \rangle$} 
&\multicolumn{2}{c}{\momn{\bt}{4}} 
&\multicolumn{2}{c}{\momn{\bw}{4}} 
&\multicolumn{2}{c}{$\langle (\ytwothree)^4 \rangle$} 
&\multicolumn{2}{c }{\momn{\mh}{4}} 
 \\ 
 
 \noalign{\smallskip}\hline\noalign{\smallskip} 
{\bf \boldmath \asmz}&$0.1553$&&$0.1546$&&$0.1489$&&$0.1392$&&$0.1333$&&$0.1399$&
 \\ 
Statistical error&$0.0018$&&$0.0009$&&$0.0010$&&$0.0019$&&$0.0045$&&$0.0010$&
 \\ 
Experimental syst.&$0.0024$&&$0.0024$&&$0.0017$&&$0.0042$&&$0.0045$&&$0.0013$&
 \\ 
HERWIG hadr.~corr. &$+0.0051$&&$+0.0031$&&$+0.0030$&&$-0.0009$& &$+0.0034$& &$+0.0083$& 
 \\ 
ARIADNE hadr.~corr.&$+0.0022$& &$+0.0022$& &$+0.0013$& &$-0.0068$&&$-0.0039$&&$-0.0036$&
 \\ 
Hadronisation error &$ 0.0051$& &$ 0.0031$& &$ 0.0030$& &$ 0.0068$& &$ 0.0039$& &$0.0083$&
 \\ 
$x_\mu$ variation:\\
 \hspace{.3cm}${\xmu=2.0} $&${+0.0183} $&&${+0.0185} $&&${+0.0187} $&&${+0.0083} $&&${+0.0079} $&&${+0.0094} $&\\
 \hspace{.3cm}${\xmu=0.5}$&$ {-0.0145}$&&$ {-0.0146}$&&$ {-0.0148}$&&$ {-0.0060}$&&$ {-0.0060}$&&$ {-0.0073}$&
 \\ 
\chisqd&$10.9/5$&&$17.3/5$&&$17.3/5$&&$24.4/5$&&$3.2/5$&&$47.0/5$&
 \\ 
\noalign{\smallskip}\hline\hline\noalign{\smallskip} 
&\multicolumn{2}{c}{$\langle (1-T)^5 \rangle$} 
&\multicolumn{2}{c}{$\langle C^5 \rangle$} 
&\multicolumn{2}{c}{\momn{\bt}{5}} 
&\multicolumn{2}{c}{\momn{\bw}{5}} 
&\multicolumn{2}{c}{$\langle (\ytwothree)^5 \rangle$} 
&\multicolumn{2}{c}{\momn{\mh}{5}} 
 \\ 
 
 \noalign{\smallskip}\hline\noalign{\smallskip} 
{\bf \boldmath \asmz}&$0.1580$&&$0.1586$&&$0.1525$&&$0.1397$&&$0.1314$&&$0.1411$&
 \\ 
Statistical error&$0.0027$&&$0.0011$&&$0.0015$&&$0.0035$&&$0.0070$&&$0.0013$&
 \\ 
Experimental syst.&$0.0029$&&$0.0025$&&$0.0020$&&$0.0052$&&$0.0061$&&$0.0017$&
 \\ 
HERWIG hadr.~corr. &$+0.0066$&&$+0.0044$&&$+0.0040$&&$-0.0013$& &$+0.0031$& &$+0.0094$& 
 \\ 
ARIADNE hadr.~corr.&$+0.0027$& &$+0.0029$& &$+0.0012$& &$-0.0077$&&$-0.0043$&&$-0.0040$&
 \\ 
Hadronisation error &$ 0.0066$& &$ 0.0044$& &$ 0.0040$& &$ 0.0077$& &$ 0.0043$& &$0.0094$&
 \\ 
$x_\mu$ variation: \\
\hspace{.3cm}${\xmu=2.0} $&${+0.0198} $&&${+0.0204} $&&${+0.0206} $&&${+0.0078} $&&${+0.0075} $&&${+0.0096} $&\\
\hspace{.3cm}${\xmu=0.5}$&$ {-0.0155}$&&$ {-0.0159}$&&$ {-0.0161}$&&$ {-0.0055}$&&$ {-0.0057}$&&$ {-0.0073}$&
 \\ 
\chisqd&$9.6/5$&&$18.4/5$&&$11.9/5$&&$10.5/5$&&$17.3/5$&&$32.4/5$&
 \\ \noalign{\smallskip}\hline
\end{tabular}}
\end{center}
\end{table*}

\begin{table*}
 \caption{Measurements of \asmz\  from event shape moments
   over the full analysed range of \Petra\ c.m.\ energies, 14--44\,GeV,  
  and  the full range of \Lep\ c.m.\ energies, 91--209\,GeV. 
   The hadronisation uncertainty is taken to 
   be the larger of the deviations observed using \Herwig\ and \Ariadne.
  The experimental systematic errors are estimated by the minimum overlap 
  assumption}
 \label{JOdpgminovl}
\begin{center}
\small

\resizebox{0.69\textwidth}{!}{

\begin{tabular}{  l  r@{}l r@{}l r@{}l r@{}l r@{}l r@{}l   r@{}l  }
\hline\noalign{\smallskip}                                                     
& \multicolumn{2}{c}{$\langle (1-T)^1 \rangle$}             
& \multicolumn{2}{c}{$\langle C^1 \rangle$}                 
& \multicolumn{2}{c}{\momn{\bt}{1}}                         
& \multicolumn{2}{c}{}                         
& \multicolumn{2}{c}{$\langle (\ytwothree) ^1 \rangle$}      
& \multicolumn{2}{c }{}                                     
 \\                                       
                                              
 \noalign{\smallskip}\hline\noalign{\smallskip}                                             

 {\bf \boldmath \asmz}&$0.1271$ &&$0.1242$&&$0.1165$&&&&$0.1259$&&&
 \\ 
Statistical error  &$0.0002$ &&$0.0002$&&$0.0001$&&&&$0.0005$&&&
 \\ 
Experimental syst.  &$0.0008$ &&$0.0006$&&$0.0005$&&&&$0.0016$&&&
 \\ 
HERWIG hadr.~corr. &$-0.0017$&    &$-0.0020$&    &$-0.0014$&    &&&$+0.0006$&    &&
 \\
ARIADNE hadr.~corr.&$+0.0023$&&$+0.0021$&&$+0.0019$&&&&$-0.0013$&&&
 \\ 
Hadronisation error &$ 0.0023$&    &$ 0.0021$&    & $0.0019$&    &&&$ 0.0013$&    &&
 \\ 
$x_\mu$ variation: \\
\hspace{.3cm}${\xmu=2.0} $&${+0.0078} $&&${+0.0071} $&&${+0.0053} $&&&&${+0.0060} $&&&\\
\hspace{.3cm}${\xmu=0.5}$&$ {-0.0063}$&&$ {-0.0057}$&&$ {-0.0041}$&&&&$ {-0.0045}$&&&
 \\ 
\chisqd &$31.2/9$& &$36.1/9$& &$114/9$& && &$173/9$&&&
 \\ 
\noalign{\smallskip}\hline\hline\noalign{\smallskip} 
&\multicolumn{2}{c}{$\langle (1-T)^2 \rangle$} 
&\multicolumn{2}{c}{$\langle C^2 \rangle$} 
&\multicolumn{2}{c}{\momn{\bt}{2}} 
&\multicolumn{2}{c}{\momn{\bw}{2}} 
&\multicolumn{2}{c}{$\langle (\ytwothree) ^2 \rangle$} 
&\multicolumn{2}{c }{\momn{\mh}{2}} 
 \\ 
 
 \noalign{\smallskip}\hline\noalign{\smallskip} 
{\bf \boldmath \asmz}&$0.1437$&&$0.1414$&&$0.1338$&&$0.1271$&&$0.1279$&&$0.1253$&
 \\ 
Statistical error&$0.0005$&&$0.0003$&&$0.0003$&&$0.0004$&&$0.0012$&&$0.0003$&
 \\ 
Experimental syst.&$0.0013$&&$0.0011$&&$0.0008$&&$0.0013$&&$0.0022$&&$0.0009$&
 \\ 
HERWIG hadr.~corr. &$+0.0007$& &$-0.0002$& &$-0.0002$& &$-0.0009$& &$+0.0018$&&$+0.0034$& 
 \\ 
ARIADNE hadr.~corr.&$+0.0026$&&$+0.0026$&&$+0.0019$&&$-0.0021$&&$-0.0009$&&$+0.0003$&
 \\ 
Hadronisation error &$ 0.0026$& &$ 0.0026$& &$ 0.0019$& &$ 0.0021$& &$ 0.0018$& &$ 0.0034$&
 \\ 
$x_\mu$ variation: \\
\hspace{.3cm}${\xmu=2.0} $&${+0.0131} $&&${+0.0126} $&&${+0.0122} $&&${+0.0061} $&&${+0.0066} $&&${+0.0052} $&\\
\hspace{.3cm}${\xmu=0.5}$&$ {-0.0107}$&&$ {-0.0103}$&&$ {-0.0100}$&&$ {-0.0045}$&&$ {-0.0051}$&&$ {-0.0036}$&\\
\chisqd&$22.3/9$& &$24.7/9$& &$47.0/9$& &$230/9$& &$46.4/9$& &$247/9$&
 \\ 
\noalign{\smallskip}\hline\hline\noalign{\smallskip} 
&\multicolumn{2}{c}{$\langle (1-T)^3 \rangle$} 
&\multicolumn{2}{c}{$\langle C^3 \rangle$} 
&\multicolumn{2}{c}{\momn{\bt}{3}} 
&\multicolumn{2}{c}{\momn{\bw}{3}} 
&\multicolumn{2}{c}{$\langle (\ytwothree) ^3 \rangle$} 
&\multicolumn{2}{c }{\momn{\mh}{3}} 
 \\ 
 
 \noalign{\smallskip}\hline\noalign{\smallskip} 
{\bf \boldmath \asmz}&$0.1509$&&$0.1495$&&$0.1436$&&$0.1317$&&$0.1282$&&$0.1307$&
 \\ 
Statistical error&$0.0010$&&$0.0005$&&$0.0005$&&$0.0009$&&$0.0021$&&$0.0004$&
 \\ 
Experimental syst.&$0.0016$&&$0.0014$&&$0.0011$&&$0.0020$&&$0.0025$&&$0.0012$&
 \\ 
HERWIG hadr.~corr. &$+0.0026$& &$+0.0011$& &$+0.0011$& &$-0.0008$& &$+0.0022$&&$+0.0047$& 
 \\ 
ARIADNE hadr.~corr.&$+0.0027$&&$+0.0031$&&$+0.0018$&&$-0.0034$&&$-0.0012$& &$-0.0002$&
 \\ 
Hadronisation error &$ 0.0027$& &$ 0.0031$& &$ 0.0018$& &$ 0.0034$& &$ 0.0022$& &$0.0047$&
 \\ 
$x_\mu$ variation:\\
\hspace{.3cm}${\xmu=2.0} $&${+0.0158} $&&${+0.0154} $&&${+0.0159} $&&${+0.0069} $&&${+0.0067} $&&${+0.0071} $&\\
\hspace{.3cm}${\xmu=0.5}$&$ {-0.0127}$&&$ {-0.0125}$&&$ {-0.0128}$&&$ {-0.0052}$&&$ {-0.0051}$&&$ {-0.0055}$&\\
\chisqd&$15.9/9$& &$22.0/9$& &$28.8/9$& &$117/9$& &$16.0/9$& &$194/9$&
 \\ 
\noalign{\smallskip}\hline\hline\noalign{\smallskip} 
&\multicolumn{2}{c}{$\langle (1-T)^4 \rangle$} 
&\multicolumn{2}{c}{$\langle C^4 \rangle$} 
&\multicolumn{2}{c}{\momn{\bt}{4}} 
&\multicolumn{2}{c}{\momn{\bw}{4}} 
&\multicolumn{2}{c}{$\langle (\ytwothree) ^4 \rangle$} 
&\multicolumn{2}{c }{\momn{\mh}{4}} 
 \\ 
 
 \noalign{\smallskip}\hline\noalign{\smallskip} 
{\bf \boldmath \asmz}&$0.1555$&&$0.1551$&&$0.1490$&&$0.1340$&&$0.1284$&&$0.1329$&
 \\ 
Statistical error &$0.0015$&&$0.0006$&&$0.0009$&&$0.0017$&&$0.0036$&&$0.0007$&
 \\ 
Experimental syst.&$0.0020$&&$0.0016$&&$0.0014$&&$0.0029$&&$0.0040$&&$0.0015$&
 \\ 
HERWIG hadr.~corr. &$+0.0043$&&$+0.0022$& &$+0.0023$&&$-0.0009$& &$+0.0024$&&$+0.0056$& 
 \\ 
ARIADNE hadr.~corr.&$+0.0030$& &$+0.0037$&&$+0.0016$& &$-0.0046$&&$-0.0019$& &$-0.0008$&
 \\ 
Hadronisation error &$ 0.0043$& &$ 0.0037$& &$ 0.0023$& &$ 0.0046$& &$ 0.0024$& &$0.0056$&
 \\ 
$x_\mu$ variation: \\
\hspace{.3cm}${\xmu=2.0} $&${+0.0179} $&&${+0.0177} $&&${+0.0185} $&&${+0.0070} $&&${+0.0067} $&&${+0.0076} $&\\
\hspace{.3cm}${\xmu=0.5}$&$ {-0.0142}$&&$ {-0.0141}$&&$ {-0.0146}$&&$ {-0.0051}$&&$ {-0.0051}$&&$ {-0.0058}$&\\
\chisqd&$13.0/9$& &$21.7/9$& &$20.3/9$& &$50.2/9$& &$6.6/9$& &$139/9$&
 \\ 
\noalign{\smallskip}\hline\hline\noalign{\smallskip} 
&\multicolumn{2}{c}{$\langle (1-T)^5 \rangle$} 
&\multicolumn{2}{c}{$\langle C^5 \rangle$} 
&\multicolumn{2}{c}{\momn{\bt}{5}} 
&\multicolumn{2}{c}{\momn{\bw}{5}} 
&\multicolumn{2}{c}{$\langle (\ytwothree) ^5 \rangle$} 
&\multicolumn{2}{c }{\momn{\mh}{5}} 
 \\ 
 
 \noalign{\smallskip}\hline\noalign{\smallskip} 
{\bf \boldmath \asmz}&$0.1588$&&$0.1598$&&$0.1528$&&$0.1349$&&$0.1277$&&$0.1336$&
 \\ 
Statistical error&$0.0024$&&$0.0007$&&$0.0014$&&$0.0032$&&$0.0058$&&$0.0010$&
 \\ 
Experimental syst.&$0.0029$&&$0.0016$&&$0.0019$&&$0.0044$&&$0.0065$&&$0.0021$&
 \\ 
HERWIG hadr.~corr. &$+0.0059$&&$+0.0031$& &$+0.0034$&&$-0.0012$& &$+0.0024$& &$+0.0062$& 
 \\ 
ARIADNE hadr.~corr.&$+0.0032$& &$+0.0042$&&$+0.0014$& &$-0.0056$&&$-0.0025$&&$-0.0013$&
 \\ 
Hadronisation error &$ 0.0059$& &$ 0.0042$& &$ 0.0034$& &$ 0.0056$& &$ 0.0025$& &$0.0062$&
 \\ 
$x_\mu$ variation:\\
\hspace{.3cm}${\xmu=2.0} $&${+0.0197} $&&${+0.0198} $&&${+0.0206} $&&${+0.0067} $&&${+0.0065} $&&${+0.0076} $&\\
\hspace{.3cm}${\xmu=0.5}$&$ {-0.0155}$&&$ {-0.0156}$&&$ {-0.0160}$&&$ {-0.0046}$&&$ {-0.0049}$&&$ {-0.0058}$& \\ 
\chisqd&$11.6/9$& &$23.6/9$& &$13.9/9$& &$19.0/9$& &$3.1/9$& &$93.3/9$&
 \\ \noalign{\smallskip}\hline
\end{tabular}}
\end{center}
\end{table*}
\clearpage


%
%
%
\end{document}